\documentclass[usenatbib]{mn2e}
\usepackage{latexsym,hhline,epsfig,longtable}
\usepackage{graphicx,color}  
\usepackage{ulem}

\def\chem#1{$^{#1}$}  

\title[Super and massive AGB stars yields]{Super and Massive AGB Stars: II - Nucleosynthesis and Yields - Z=0.02, 0.008 and 0.004.}
\author[C.L. Doherty, P. Gil-Pons, H.H.B Lau, J.C. Lattanzio, L. Siess]{Carolyn L. Doherty$^{1}$\thanks{E-mail:carolyn.doherty@monash.edu}, Pilar Gil-Pons$^{2,1}$, Herbert H.B Lau$^{3,1}$, John C. Lattanzio$^{1}$ 
\newauthor and Lionel Siess$^{4}$\\ 
 $^{1}$Monash Centre for Astrophysics (MoCA), School of 
 Mathematical Sciences, Monash University, Victoria 3800, Australia\\
 $^{2}$Department of Applied Physics, Polytechnical University of Catalonia, 08860 Barcelona, Spain\\
 $^{3}$Argelander Institute for Astronomy, University of Bonn, Auf dem Huegel 71, D-53121 Bonn, Germany\\
 $^{4}$Institut d'Astronomie et d'Astrophysique, Universit\'e Libre de
 Bruxelles (ULB), CP 226, B-1050 Brussels, Belgium} 

\begin{document}

 \maketitle

 \begin{abstract}

We have computed detailed evolution and nucleosynthesis models for super and massive AGB stars over the mass range 6.5 - 9.0 M$_\odot$ in divisions of 0.5 M$_\odot$ with metallicities Z=0.02, 0.008 and 0.004. These calculations, in which we find third dredge-up and hot bottom burning, fill the gap between existing low and intermediate-mass AGB star models and high mass star models that become supernovae.  
For the considered metallicities, the composition  of the yields is largely dominated by the thermodynamic conditions at the base of the convective envelope rather than by the pollution arising from third dredge up. 
We investigate the effects of various uncertainties, related to the mass-loss rate, mixing length parameter, and the treatment of evolution after the envelope instability that develops near the end of the (Super)AGB phase. Varying these parameters alter the yields mainly because of their impact on the amount of third dredge up enrichment, and to a lesser extent on the hot bottom burning conditions. 
Our models produce significant amounts of \chem{4}He, \chem{7}Li (depending on the mass-loss formulation) \chem{13}C, \chem{14}N, \chem{17}O, \chem{23}Na, \chem{25}Mg, as well the radioactive isotope \chem{26}Al in agreement with previous investigation. In addition our results show enrichment of \chem{22}Ne, \chem{26}Mg and \chem{60}Fe, as well as a substantial increase in our proxy neutron capture species representing all species heavier than iron.
These stars may provide important contributions to the Galaxy's inventory of the heavier Mg isotopes, \chem{14}N, \chem{7}Li and \chem{27}Al.

\end{abstract}


\section{Introduction}

Super AGB stars occupy a narrow initial mass range between $\approx$ 6.5-10 M$_\odot$ depending on composition. Due to the shape of the initial mass function (IMF) they are nevertheless quite numerous and may contribute substantially to the galactic inventory of certain isotopes. 
This mass range has been largely understudied due to two main reasons. Firstly, the difficulties in accurately following the partially degenerate carbon burning phase of evolution.  Secondly, they are also computationally demanding, with the thermally pulsing phase consisting of hundreds or even thousands of thermal pulses. 

Whilst evolutionary models for solar metallicity super AGB stars during core carbon burning and the first few thermal pulses have existed for almost 20 years \citep{pap1}, only recently have studies looked into detailed nucleosynthesis and stellar yields.
\cite{sie10} produced a grid of nucleosynthetic yields for the metallicity range Z=0.04-10$^{-4}$, whilst earlier work looked at specific isotopes such as \chem{26}Al \citep{sie08b}. 
A selection of works \cite[e.g.][]{ven10a,ven11b,ven13,her12} have explored nucleosynthetic yields of Pop II super AGB stars as prospective polluter candidates to explain the globular cluster multiple populations problem.  
Another area of super AGB nucleosynthesis research to date has been the study of s-process isotope production at solar metallicity \cite[e.g.][]{lau11,kar12}.
Whilst these calculations represent an important advance in our knowledge of these stars, we still lack a comprehensive study over a range of masses which self consistently includes the effects of third dredge-up (hereafter 3DU).

We outline three possible uses for detailed nucleosynthesis predictions from models of super and massive AGB stars. 
Firstly by comparing nucleosynthetic surface abundances to direct observations of massive AGB stars we may be able to constrain mixing processes or mass-loss rates.
Secondly, these nucleosynthetic and yield calculations will be of benefit to the galactic chemical evolution (GCE) community for inclusion in their models. 
Since the beginnings of GCE modelling, this mass range has been omitted, due to the lack of computed models. This ``super AGB star gap'' in stellar yield inputs is most strikingly seen graphically in the works of \cite{rom10a} and \cite{kob11}.
The mass boundary between stars which end their lives as white dwarfs or supernovae is under considerable study \cite[e.g.][]{sie07,poe08,sma09a}. It is important from a GCE point of view as low mass supernova and high mass AGB stars vastly differ in their element production.
Thirdly, of great interest is the possibility of a unique nucleosynthetic signature which would differentiate a super AGB star from a massive AGB star.

In this paper we present the first full thermally pulsing super AGB yields calculations over a wide range of metallicities taking into account the effects of both hot bottom burning (HBB) and 3DU.
This paper is number II in this series on super and massive AGB stars. Paper I \citep{doh10} made detailed comparisons of carbon burning behaviour between different stellar evolution codes. Papers III, IV and V (Doherty et al. 2013, in preparation) explore a) nucleosynthetic yields at globular cluster metallicities, b) the final fates of these objects, and c) the efficiency of third dredge-up at various metallicities.
In this paper, Section 2 covers the numerical programs and describes the input physics and assumptions. 
In Section 3 we shall discuss the nucleosynthesis, in Section 4 we present a suite of detailed nucleosynthetic yields. Finally in Section 5 we summarize and draw the main conclusions from our work.

\section{Numerical Programs}\label{sec-np}

The detailed stellar evolution was calculated using the Monash University stellar evolution program (\texttt{MONSTAR}) as described in \cite{fro96}, \cite{cam08} and \cite{doh10}. The relevant input physics can be summarised as follows.
 We use the \cite{rei75} mass-loss rate with $\eta$=1.0 following \cite{blo95b} for all phases of evolution prior to carbon burning/or AGB phase. This results in between $\sim$ 0.1-0.5 M$_\odot$ mass lost prior to the first thermal pulse. We then use our standard mass-loss rate from \cite{vas93}. For the mixing length parameter we use $\alpha_{\rm{mlt}}$=1.75, calibrated to match the present day value of the solar radius. We employ the search for convective neutrality as described in \cite{lat86}. Mid temperature opacities are from the OPAL compilation \citep{opal} whilst low temperature opacities are from \cite{fer05b}. In the calculation of the gravo-thermal energy production rate, we take into account the \cite{woo81} entropy of mixing correction.

The calculation of hundreds of thermal pulses was computationally demanding, with in excess of 20 million evolutionary time steps required for the most massive models. The temporal resolution required was very fine so we could follow the extraordinary short thermal pulse duration (less than 3 months!). The mesh resolution was also very fine to accurately model the 10$^{-4}$ M$_\odot$ region that contained not only both the H and He burning shells but also the base of the convective envelope (BCE). Our evolutionary models generally have between 3000-5000 mass shells and time-steps down to 10$^{-4}$ years. The longest running evolution calculation took of the order of 3 months CPU time on a single 2.2 GHz processor desktop computer, whilst depending on mass and metallicity of model typically we calculate between 10-15 thermal pulses per day. We have computed a grid of super and massive AGB stars in 0.5 M$_\odot$ initial mass divisions.

Nucleosynthesis calculations were performed using the Monash stellar nucleosynthesis post processing program \texttt{MONSOON}. A post-processing code offers many advantages for the calculations presented here, primarily computational expediency, but at some cost in versatility. 

We calculate the structure of the star using \texttt{MONSTAR}, and then send the details of the structure (temperatures, densities, radii, mixing lengths and convective velocities, but not compositions) to \texttt{MONSOON} to use as the basis on which it calculates detailed nuclear reactions and mixing.
The advantage is that a different spatial mesh and time-stepping algorithm can be used for the nucleosynthesis. The evolution code is slowest during the interpulse phase because of the accuracy requirements presented by the movement (in mass) of the nuclear shells. This is accomplished quickly in \texttt{MONSOON} by reading in the positions (in mass) of the nuclear shells. \texttt{MONSOON} takes more time during the thermal pulses themselves, when much nucleosynthesis occurs, but when the structure is changing relatively slowly. We employ a ``donor cell'' advective scheme with two stream (up and down) mixing, and solve simultaneously the chemical transport and nuclear burning \cite[for details see][]{can93,lug04}.

The disadvantage of this approach is that changes in the composition do not feed back on the structure, which was already calculated using \texttt{MONSTAR}. Changes in composition have three effects on the structure, specifically they alter i) the mean molecular weight (and hence the equation of state), ii) the opacity and iii) the energy generation. However, all major species have been included in the structure calculation, and the minor species included here (Li, Ne, Na, Mg etc) are at such low concentrations that changes in their abundance have an entirely negligible effect on the structure. Hence in this case it is quite safe to use a post-processing code.

The nuclear network comprises 77 species up to sulfur (including both ground and metastable states of \chem{26}Al), as well as the iron group elements, using the double sink approach as originally proposed by \cite{jor89}. To terminate the network and access the level of neutron capture reactions past nickel, we use a fictitious particle $g$ \citep{lug04} which is our s-process proxy. The initial abundances for the considered species were scaled with metallicity from the solar composition of \cite{gre96} as given in Appendix B1.  

Typically not every time-step from \texttt{MONSTAR} is necessary for the post-processing code. We output evolutionary models to be used as input for \texttt{MONSOON} according to an algorithm that checks for changes in the stellar structure and evolutionary stage. \texttt{MONSTAR} writes a model for \texttt{MONSOON} according to an algorithm that specifies maximum changes in the internal structure between successive outputs. These can be at every timestep during a thermal pulse/third dredge up event, but are every 10-50 timesteps during the interpulse phase. Also due to the differing meshing within the nucleosynthesis code only about 800-1000 mass shells are required (each with an up and downstream composition).

We have refined the post-processing meshing routine to be suitable for (S)AGB models, and now include the central regions of the star so that we can follow the core composition, noting however that this has no effect on the yields. 

The majority of the reaction rates are from the JINA reaction library of \cite{jin10}. The proton capture reactions for the neon-sodium (Ne-Na) and magnesium-aluminium (Mg-Al) cycles are from \cite{ili01}. We use the \cite{kar06a} rate for the \chem{22}Ne($\alpha$,n)\chem{25}Mg and \chem{22}Ne($\alpha,\gamma$)\chem{26}Mg reactions and the  \chem{23}Na(p,$\gamma$)\chem{24}Mg and \chem{23}Na(p,$\alpha$)\chem{20}Ne from \cite{hal04} and \chem{22}Ne(p,$\gamma$)\chem{23}Na from \cite{hal02}.

\section{Nucleosynthesis}
For detailed description of the thermally pulsing phase of evolution for super AGB stars
refer to \cite{sie10} and Paper IV (Doherty et al. 2013 in preparation) in this series. In this section we consider the four events that affect the surface composition: first, second and third dredge-up events and hot bottom burning. 

\subsection{First and Second Dredge-up}

After the cessation of core hydrogen burning, intermediate-mass stars of solar and moderate metallicity climb the RGB. This structural readjustment leads to the occurrence of the first dredge-up (1DU), when the convective envelope penetrates into the region that has previously undergone hydrogen burning.
During 1DU the surface abundances of  H, \chem{7}Li, \chem{12}C, \chem{15}N, \chem{18}O show quite a significant decrease, whilst to a lesser extent the surface abundance of \chem{16}O, \chem{25}Mg, \chem{22}Ne and \chem{19}F also decrease.
 Acting as a counterbalance \chem{3}He, \chem{4}He, \chem{14}N, \chem{13}C, \chem{17}O are greatly enriched at the surface with slight increases in \chem{26}Mg, \chem{23}Na and \chem{21}Ne.
All of these changes are to be expected with the dredge-up of material from regions of partial hydrogen burning in which the CNO cycle is fully activated but with only partial activation of the heavier hydrogen burning cycles/chains.

The second dredge-up (2DU) occurs in intermediate-mass stars after core helium burning during the ascent of the giant branch. For most of the models here, 2DU proceeds as in intermediate-mass AGB stars, and causes changes in species involved in H burning which are observed at the surface. The most massive models in this work undergo corrosive 2DU, where the base of the convective envelope dips into the He burning shell which results in an increase in the surface abundance of He burning products, mostly \chem{12}C.
After 2DU, the C/O ratio is less than unity in all models apart from the 8 M$_\odot$ Z=0.004 (Fig.~\ref{fig-coratio}).  
There is a very large enhancement of \chem{4}He with up to $\Delta$Y$\sim$0.1. 

The initial, 1DU and 2DU surface abundances (in mass fraction) for our standard models can be found in Table~\ref{12du} in Appendix B. We note close agreement with \cite{sie07}.

\subsection{Thermally Pulsing Super and Massive AGB phase}\label{sec-tp}

\begin{figure}
\resizebox{\hsize}{!}{\includegraphics{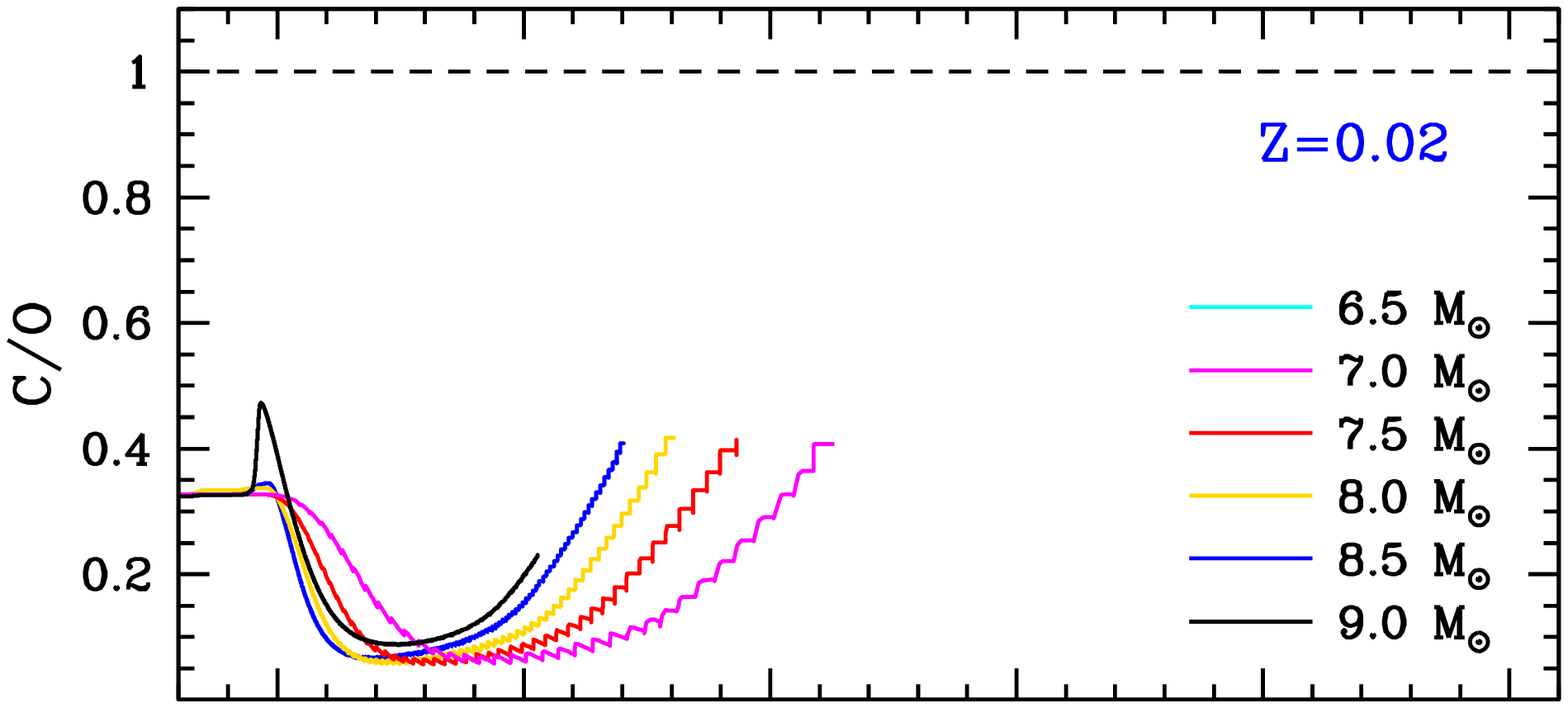}}
\resizebox{\hsize}{!}{\includegraphics{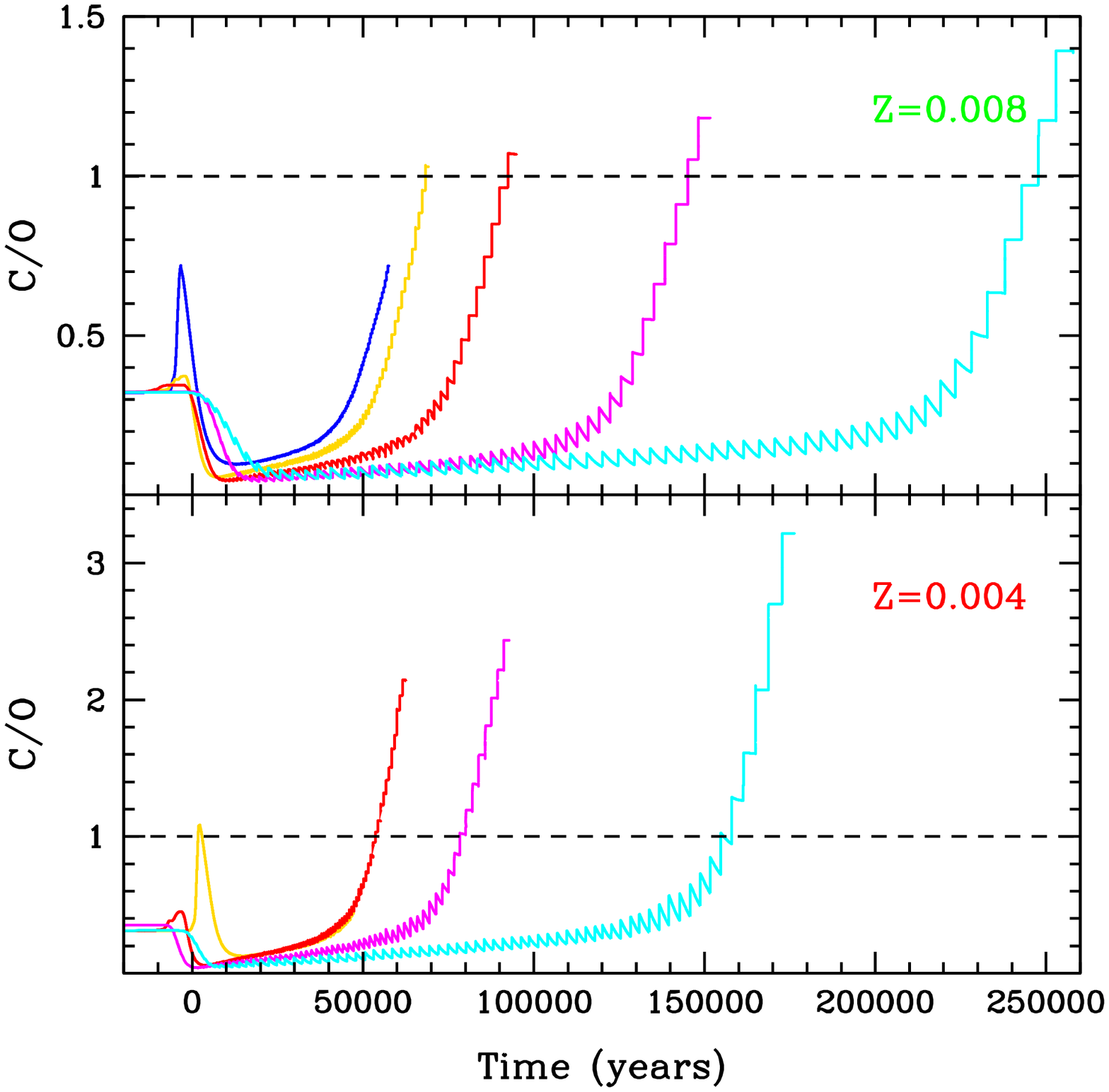}}
\caption{Surface abundances of the number ratio C/O for our standard VW93 mass-loss rates cases. The horizontal dashed line represents the divide between M stars (O-rich) and the C stars (C-rich). The time axis has been offset with the zero at the time of first thermal pulse.}\label{fig-coratio} 
\end{figure}

In this section we explore the thermally pulsing phase of evolution of super and massive AGB stars. 
Due to their more massive and compact cores the thermal pulses are very short lived (of the order of a year) and the region which contains the H and He shells and intershell is very thin in mass
($\sim$10$^{-5}$ M$_\odot$). The interpulse periods are also correspondingly shorter, decreasing with increasing initial mass. Due to these factors very fine spatial and temporal resolution is vital to accurately model the TP-(S)AGB stars.
In the mass and metallicity range covered here, all but the lowest mass, and most metal poor models, stay oxygen rich for the majority of their evolution. In Figure~\ref{fig-coratio} we show the variation of the C/O (number) ratio with time for our standard models.

Inclusion of low temperature variable composition molecular opacities has been shown to be important in AGB models when the C/O ratio exceeds unity. When this value is reached,  molecule formation drives a large increase in opacity, a drop in the effective temperature and a subsequent enhancement in the mass-loss rate \cite[e.g.][]{mar02,cri07}. We have not included these opacities in the work presented here. We believe that this is not a serious weakness in our case because all of our standard models become carbon rich only after the star has entered the super-wind phase. Thus the mass-loss rate is already very high and the evolution is rapidly terminated even without the extra opacity from carbon-rich molecules. We note that the new envelope opacities are included in our current version of \texttt{MONSTAR}, but not the one used for the calculations presented in this paper.

Table~\ref{tpsagb} contains structural information of most relevance to the nucleosynthesis. Further variables such as the maximum quiescent luminosity during the TP-(S)AGB phase ($L^{\rm{Max}}$), C/O number ratio at the completion of 2DU and the end of the evolution, and the ratio of carbon rich to oxygen rich duration of the TP-(S)AGB phase ($\tau_{\rm{C}}$/$\tau_{\rm{M}}$) can be found in Table~\ref{extrapstuff} in Appendix B. A more detailed description of these evolutionary models can be found in Paper IV (Doherty et al. 2013, in preparation) in this series. The M labeled rows are for the delayed super-wind mass-loss rate case that shall be discussed in Section~\ref{sec-ml}.
During the TP-(S)AGB phase of evolution, nucleosynthesis takes place in three main regions; the hydrogen burning shell (HBS), the base of the convective envelope (via HBB) and during a thermal pulse in the intershell between the hydrogen and helium burning shells (HeBS). The evolution of the surface abundance is determined primarily by the competing effects of hot bottom burning and third dredge-up.

\begin{table}
\begin{center}
\setlength{\tabcolsep}{1.5pt} 
\caption{Selected model characteristics. $T_{\rm{BCE}}^{\rm{Max}}$ is the maximum temperature at the base of the convective envelope; $T_{\rm{He}}$ is the maximum temperature in the helium burning intershell region; $M_{\rm{Dredge}}^{\rm{Tot}}$ is the total mass of material dredged-up to the surface due to 3DU; $M_{\rm{2DU}}$ is the post 2DU core mass; $M_{\rm{C}}^{\rm{F}}$ is the final computed core mass; $M_{\rm{env}}^{\rm{F}}$ is the final envelope mass, $N_{\rm{TP}}$ is the number of thermal pulses and $\tau_{\rm{(S)AGB}}$ is the TP-(S)AGB lifetime. The M indicates our modified VW93 mass-loss rate. The values in square brackets represent the number of further extrapolated thermal pulses. (Note that $n(m)= n\times 10^{m}$.) For details of extrapolated thermal pulse characteristics refer to Section~\ref{ref-extrap} and Table~\ref{extrapstuff}.}
\label{tpsagb}
\begin{tabular}{lccccccrc} \hline \hline 
$M_{\rm{ini}}$&$T_{\rm{BCE}}^{\rm{Max}}$&$T_{\rm{He}}$& $M_{\rm{Dredge}}^{\rm{Tot}}$ &$M_{\rm{2DU}}$& $M_{\rm{C}}^{\rm{F}}$ &$M_{\rm{env}}^{\rm{F}}$ &  $N_{\rm{TP}}$ & $\tau_{\rm{(S)AGB}}$ \\
(M$_\odot$)&(MK)&(MK)&(M$_\odot$)&(M$_\odot$)&(M$_\odot$)&(M$_\odot$)&&(yrs) \\ 
\hline \multicolumn{9}{c}{Z=0.02}   \\ \hline
7.0  &84 & 374 &4.45(-2)&0.96&0.97&1.54 &38[7]   & 1.11(5)  \\
7.0M &89 & 401 &5.00(-1)&0.95&1.01&1.65 &321[12] & 1.00(6)  \\
7.5  &88 & 383 &3.86(-2)&1.00&1.01&1.64 &45[13]  & 9.14(4)  \\
7.5M &93 & 404 &2.75(-1)&1.00&1.04&1.60 &264[14] & 5.09(5)  \\
8.0  &97 & 402 &3.82(-2)&1.06&1.07&1.61 &62[16]  & 7.78(4)  \\
8.0M &100& 409 &1.31(-1)&1.06&1.09&1.74 &193[21] & 2.39(5)  \\
8.5  &106& 414 &3.64(-2)&1.14&1.15&1.78 &110[35] & 6.60(4)  \\
8.5M &106& 416 &4.75(-2)&1.14&1.16&1.87 &149[40] & 8.63(4)  \\
9.0  &113& 419 &3.53(-2)&1.21&1.23&1.82 &221[76] & 6.39(4)  \\
\hline \multicolumn{9}{c}{Z=0.008} \\ \hline
6.5  &90 & 385 &1.07(-1)&0.94 &0.95&0.95& 63[5]  & 2.58(5)  \\ 
6.5M &90 & 405 &4.88(-1)&0.94 &0.99&0.97&283[5]  & 1.04(6)  \\
7.0  &95 & 387 &6.38(-2)&0.99 &1.00&1.05& 56[7]  & 1.49(5)  \\ 
7.0M &95 & 411 &4.18(-1)&0.99 &1.04&1.13&356[12] & 8.13(5)  \\
7.5  &99 & 400 &4.18(-2)&1.05 &1.06&1.14& 58[9]  & 8.95(4)  \\ 
7.5M &102& 418 &2.78(-1)&1.05 &1.09&1.25&465[15] & 6.62(5)  \\
8.0  &110& 411 &3.44(-2)&1.13 &1.14&1.21& 93[21] & 6.60(4)  \\ 
8.0M &110& 418 &2.00(-1)&1.13 &1.19&1.63&690[46] & 3.71(5)  \\
8.5  &117& 418 &3.11(-2)&1.21 &1.22&1.42&185[53] & 5.84(4)  \\ 
8.5M &118& 426 &8.06(-2)&1.21 &1.24&1.59&601[103]& 1.54(5)  \\
\hline \multicolumn{9}{c}{Z=0.004} \\ \hline
6.5  &98 & 400 &7.79(-2)&1.00&1.01&0.83 & 68[5]  & 1.73(5)  \\ 
6.5M &98 & 412 &2.08(-1)&0.99&1.02&1.01 &170[7]  & 4.28(5)  \\
7.0  &104& 394 &4.32(-2)&1.05&1.06&0.86 &65[8]   & 9.33(4)  \\ 
7.0M &104& 414 &1.70(-1)&1.05&1.08&0.94 &238[7]  & 3.27(5)  \\
7.5  &114& 410 &3.11(-2)&1.15&1.16&1.03 &104[21] & 6.16(4)  \\ 
7.5M &115& 422 &1.32(-1)&1.14&1.17&1.02 &437[23] & 2.45(5)  \\
8.0  &120& 418 &2.74(-2)&1.22&1.23&1.19 &161[78] & 5.35(4)  \\ 
8.0M &122& 416 &9.75(-2)&1.22&1.26&2.09 &955[146]& 1.89(5)  \\
\hline
\end{tabular}
\medskip\\
\end{center}
\end{table}

\subsubsection{Thermal Pulse nucleosynthesis and Third Dredge-up }\label{3duns}

Here we describe the helium burning nucleosynthesis that takes place during a thermal pulse, and the mixing of these newly generated products to the surface via third dredge-up.
The intershell region prior to a TP is composed of the ashes from the outward moving HBS and is predominately \chem{4}He, \chem{14}N and \chem{20}Ne.

\begin{figure}
\resizebox{\hsize}{!}{\includegraphics{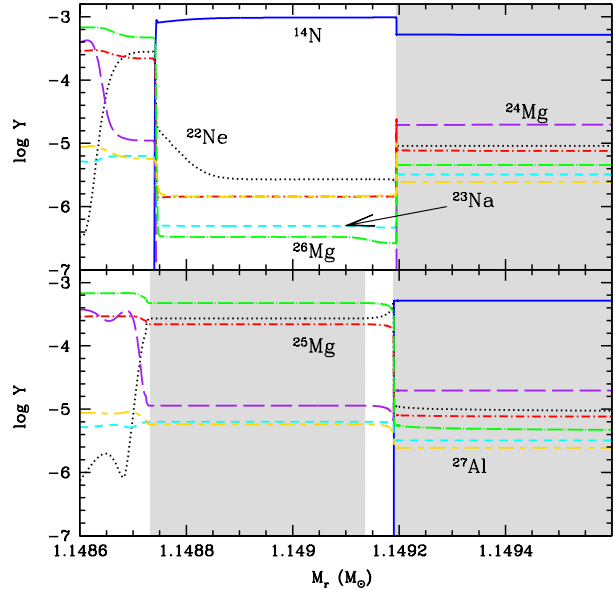}}
\caption{Composition profiles for the 8.5 M$_\odot$ Z=0.02 model during the
  50th thermal pulse. The top panel highlights the composition of the HBS, just prior to a thermal pulse whilst the bottom panel is at time $\sim$ 3 years later during the end phase of the convective thermal pulse. Grey shading represents convective regions. We do not show \chem{4}He, \chem{12}C, \chem{16}O, \chem{20}Ne so as to highlight the less abundant isotopes.}
\label{shellns}
\end{figure}

The maximum temperature within the helium burning intershell convective zone $T_{\mathrm{He}}$ steadily increases throughout the evolution. It also increases with increasing initial mass, with the most massive models in this study achieving temperatures in excess of 420MK.
 During helium burning the main energetically important reaction is clearly the triple $\alpha$. 
However, during a TP there is also a substantial production of the heavier $\alpha$ elements via \chem{12}C($\alpha,\gamma$)\chem{16}O, \chem{16}O($\alpha,\gamma$)\chem{20}Ne, \chem{20}Ne($\alpha,\gamma$)\chem{24}Mg, and even \chem{24}Mg($\alpha,\gamma$)\chem{28}Si.
Apart from the primary \chem{12}C($\alpha,\gamma$)\chem{16}O channel \chem{16}O increases with contributions from \chem{13}C($\alpha$,n)\chem{16}O and \chem{13}N($\alpha$,p)\chem{16}O.
The abundant \chem{14}N undergoes the following reactions \chem{14}N($\alpha,\gamma$)\chem{18}F($\beta^+,\nu$)\chem{18}O($\alpha,\gamma$)\chem{22}Ne. The other isotopes of neon are also produced during a TP, with \chem{20}Ne via \chem{16}O($\alpha,\gamma$)\chem{20}Ne and \chem{16}O(n,$\gamma$)\chem{17}O($\alpha$,n)\chem{20}Ne, and \chem{21}Ne via \chem{16}O(n,$\gamma$)\chem{17}O($\alpha,\gamma$)\chem{21}Ne, \chem{16}O($\alpha$,n)\chem{21}Ne and \chem{20}Ne(n,$\gamma$)\chem{21}Ne.

In these super AGB stars due to the large temperature in the helium burning intershell convective zone, $T_{\mathrm{He}}$, the \chem{22}Ne($\alpha$,n)\chem{25}Mg and \chem{22}Ne($\alpha,\gamma$)\chem{26}Mg reactions are also activated. These \chem{22}Ne nuclear reaction rates are frequently under investigation \cite[e.g.][]{nacre,kar06a,lon12} due to their importance, primarily for s-process nucleosynthesis (including the weak s-process component in massive stars), but also for the production of heavy magnesium isotopes. Above $\approx$ 330 MK, \chem{22}Ne($\alpha$,n)\chem{25}Mg is the dominant of these two reactions. 
Due to the high $T_{\mathrm{He}}$ we find a substantial neutron flux from \chem{22}Ne($\alpha$,n)\chem{25}Mg of up to 10$^{14-15}$ n/cm$^{3}$. 
Within the intershell region however, there are quite a few potent neutron poisons such as \chem{26}Al, \chem{14}N as well as \chem{25}Mg itself which absorbs many of the free neutrons via \chem{25}Mg(n,$\gamma$)\chem{26}Mg. 

After a TP the intershell region is composed of primarily $\sim$ 65-70 per cent \chem{4}He, 25-30 per cent \chem{12}C and the rest largely \chem{16}O, \chem{20}Ne, \chem{22}Ne and the heavy magnesium isotopes \chem{25,26}Mg.
After each thermal pulse, third dredge-up occurs when the convective envelope penetrates through the hydrogen burning shell and into the intershell region, mixing up products of both hydrogen and partial helium burning.
To measure the efficiency of 3DU we use the efficiency parameter\footnote{$\lambda = \Delta M_{\rm{dredge}}$/$\Delta$ $M_{\rm{H}}$, where $\Delta M_{\rm{H}}$ is the increase in the core mass during the interpulse phase and $\Delta M_{\rm{dredge}}$ is the mass of dredged-up material.} $\lambda$. 
For a full and detailed discussion of 3DU efficiency and observational evidence for the occurrence of 3DU in super and massive AGB stars see Paper V (Doherty et al. 2013 in preparation) in this series. Here we simply mention that we find efficient dredge-up in all of these super and massive AGB models, with a maximum $\lambda$ between 0.6 and 0.95. 

In Table~\ref{tpsagb} we provide the number of thermal pulses and the total amount of dredged-up material $M_{\rm{Dredge}}^{\rm{Tot}}$. Whilst the number of thermal pulses increases with increasing mass, the mass of dredged-up material, which is reliant on the thickness of the intershell, is greatly reduced. This results in $M_{\rm{Dredge}}^{\rm{Tot}}$ decreasing with increasing mass. The efficiency of 3DU also decreases with increasing mass.
As seen in previous studies of massive and/or low metallicity AGB stars, the temperature at the base of the envelope during the 3DU can remain high enough for nuclear burning to be still very active \cite[e.g.][]{chi01,sie02,gor04,her04a,lau09}. These hot third dredge-ups are not expected to significantly alter the envelope nucleosynthesis but they may suppress the formation of a \chem{13}C pocket \citep{gor04} which would then impact on the production of s-process elements. We do not include a \chem{13}C pocket in the models presented in this series of papers.

\begin{figure}
\resizebox{\hsize}{!}{\includegraphics{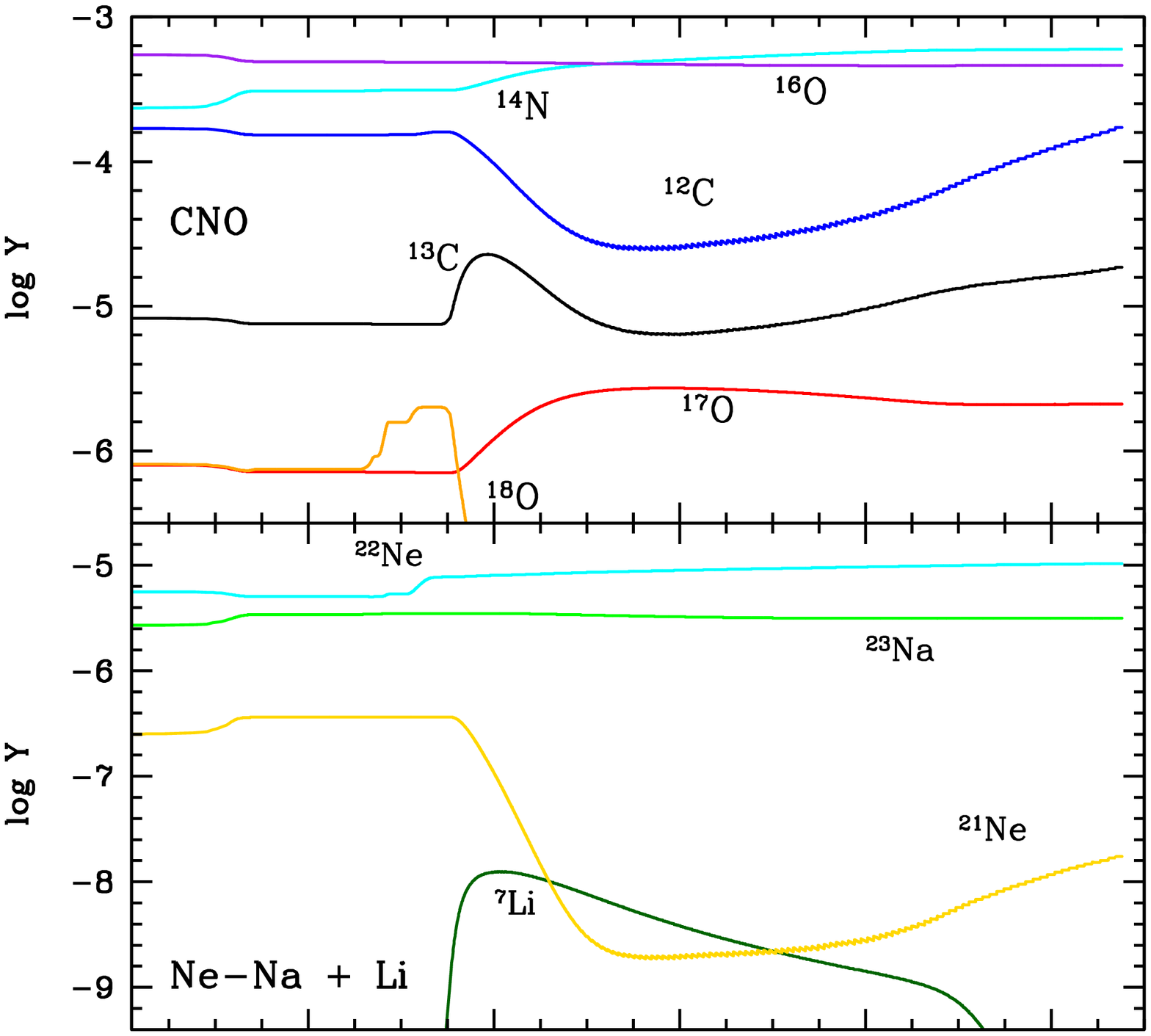}}
\resizebox{\hsize}{!}{\includegraphics{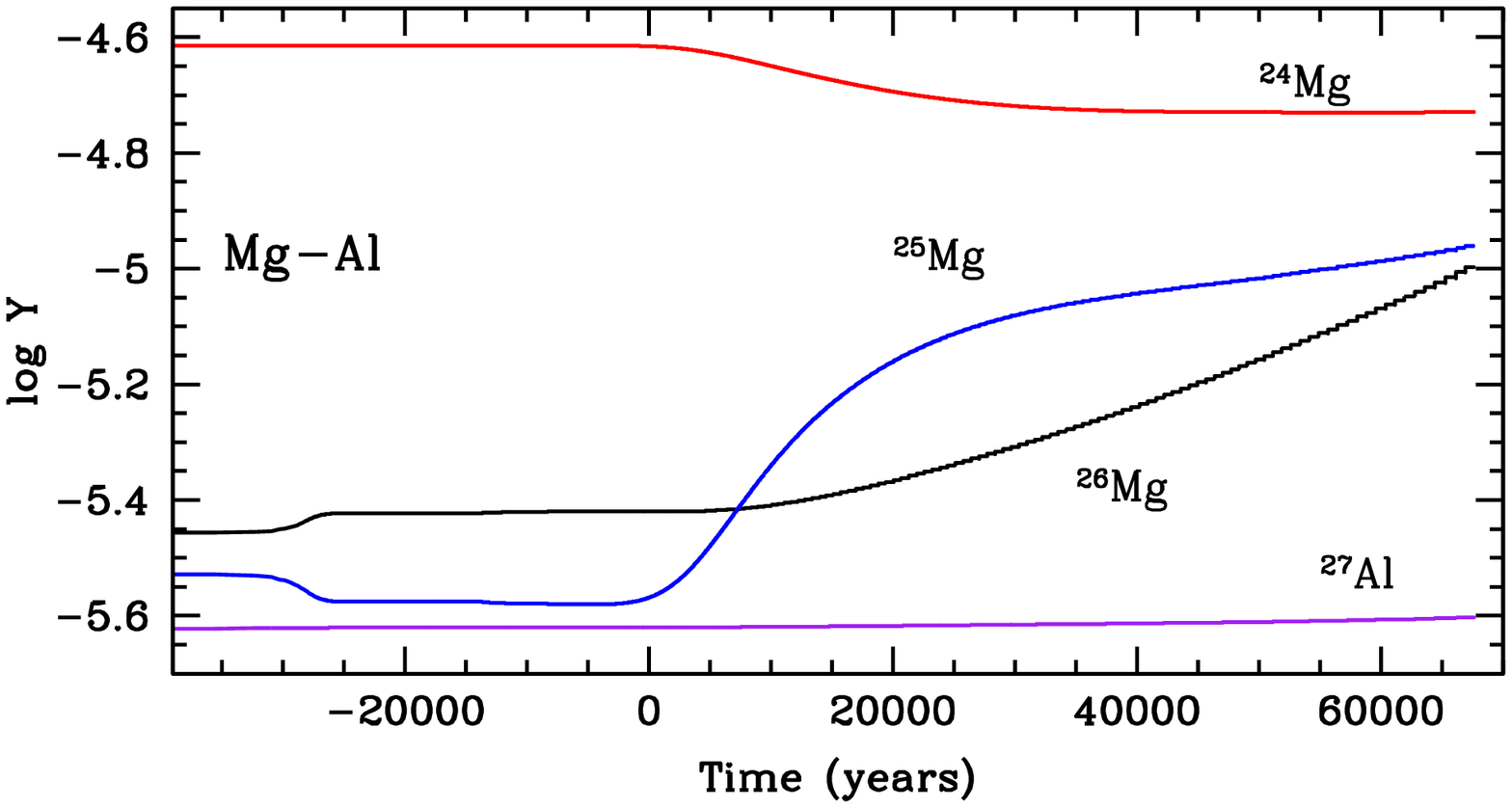}}
\caption{Surface abundances  (in logarithm of mole fraction Y) of selected isotopes as a function of time for the 8.5 M$_\odot$ Z=0.02 model. The time axis has been offset with the zero at the time of first thermal pulse.}\label{surf}
\end{figure}

\subsubsection{Hot Bottom Burning}\label{hbbns}

Hot bottom burning processes material in the thin region ($\approx$ 10$^{-5}$ M$_\odot$) at the base of the convective envelope (BCE) and in these models $T_{\rm{BCE}}$ can be as high as $\sim$ 120MK (see Table~\ref{tpsagb}).
With increasing $T_{\rm{BCE}}$, different burning cycles activate.
As the temperature reaches $\approx$ 20 MK the CNO cycle starts to operate, whilst at even higher temperatures ($\approx$ 35 MK) the Ne-Na proton capture reaction cycle activates.
When the $T_{\rm{BCE}}$, exceeds $\approx$ 30 MK \citep{sac92a} lithium is created in the convective envelope via the \cite{cam71} mechanism whereby \chem{7}Be created from \chem{4}He(\chem{3}He,$\gamma$)\chem{7}Be is quickly mixed to a cooler region where it undergoes an electron capture reaction forming \chem{7}Li. This \chem{7}Li production continues until \chem{3}He exhaustion. 
At still higher temperatures ($\approx$ 60 MK) the Mg-Al chain activates. A thorough exploration
of the effect of reaction rate uncertainties of the Ne-Na and Mg-Al
reactions in massive AGB stars can be found in \cite{izz07}. The observational evidence to support the occurrence of HBB in super and massive AGB stars relies on low \chem{12}C/\chem{13}C ratios due to CN cycling \citep{woo83}, lithium rich O-rich luminous AGB stars \cite[][]{smi90,ple93,gar07}, as well as on deviations from the core mass luminosity relation \citep{blo91}. \cite{mcs07} showed conclusive evidence for the competing processes of  HBB and 3DU in intermediate-mass AGB stars. The higher luminosity of super AGB stars may be a possible , but not very promising discriminant to distinguish them from massive AGB stars (e.g see Table~\ref{extrapstuff} in Appendix B.)
Near the end of evolution, when the envelope mass reduces below a critical value the temperature drops below that required to undergo HBB \citep{ren81}.
In all the models presented here, 3DU continues after HBB has ceased, which results in the surface composition becoming more and more enriched in carbon. However as the amount of dredged-up material decreases with increasing mass, all but the lowest mass stars and/or slowest mass-loss rate models remain O-rich for the majority of their evolution. 
This continuation of 3DU after the cessation of HBB led \cite{fro98a} to suggest this to explain the existence of luminous carbon stars \citep{van99}.

\subsubsection{Representative super AGB model: 8.5 M$_\odot$ Z=0.02}

Figure~\ref{shellns} highlights the internal nucleosynthesis taking place in the nuclearly active central regions prior to and during the 50th thermal pulse of the 8.5 M$_\odot$ Z=0.02 model. We see an exterior convective envelope in both panels, with the bottom panel also showing a convective thermal pulse region. In the HBS, the CNO isotopes have been cycled to result in a considerable amount of \chem{14}N (and \chem{4}He).
The bottom panel shows a time nearing the end of the thermal pulse where the \chem{14}N has been converted initially to \chem{22}Ne then \chem{25,26}Mg. 
Figure~\ref{surf} clearly shows the major surface abundance features caused by 3DU and HBB for our representative model. The time axis has been offset with the zero at the time of first thermal pulse.
The top panel shows the evolution of the CNO species during the 2DU and thermally pulsing phase. The \chem{12}C decreases quite rapidly at the onset of HBB as it is transformed to \chem{13}C with these two isotopes quickly coming into equilibrium. The \chem{12}C then increases during the further evolution due to 3DU. There is a slight decrease in the \chem{16}O from ON cycling.
The middle panel includes a selection of Ne and Na isotopes and \chem{7}Li. The \chem{7}Li peaks around log$\epsilon$(\chem{7}Li)\footnote{log$\epsilon$(\chem{7}Li) = log (n[Li]/n[H]) + 12, where n is number abundance} $\sim$ 4.31 before \chem{3}He is fully depleted. There is only a very slight reduction in \chem{23}Na due to HBB.
The bottom panel illustrates the changes in the surface composition of the Mg and Al isotopes. 
When the HBB temperatures exceed $\approx$ 90MK the \chem{24}Mg is efficiently depleted via proton capture reactions and converted to \chem{25}Mg. This destruction occurs quite rapidly within about 40000 years.  The \chem{25}Mg and \chem{26}Mg then increase as they are dredged to the surface.
 The \chem{4}He content (not shown) rises from its post 1DU mass fraction value of 0.287 to 0.366 due to 2DU and then further increases to 0.370 as a result of both 3DU and HBB during the thermally pulsing phase.

\section{Yields}

\begin{figure*}
\resizebox{\hsize}{!}{\includegraphics{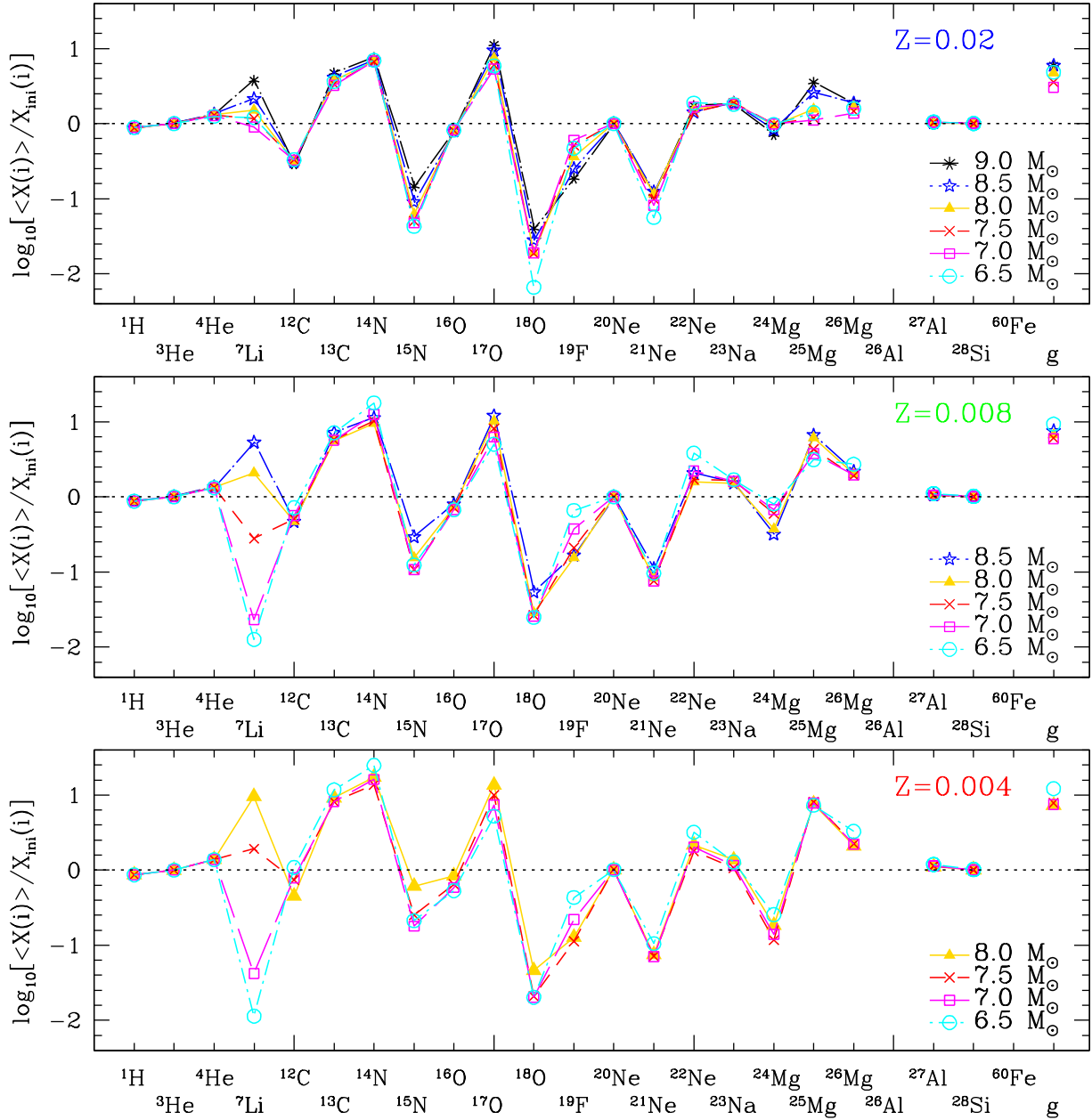}}
\caption{Production factor $\log_{10}[\langle X(i)  \rangle/X_{\mathrm{ini}}(i)]$ of selected species where $\langle X(i)   \rangle$ is the average surface mass fraction of species $i$, whilst   $X_{\mathrm{ini}}(i)$ is the initial surface mass fraction. As described in Section~\ref{sprocess} $g$ is used as a proxy for elements heavier than iron. The data for the 6.5 M$_\odot$ Z=0.02 model are from Karakas (2010).}
\label{productionfactor}
\end{figure*}

Nucleosynthetic yields (in M$_\odot$) are calculated using the following expression

\begin{equation}
 M_{i} = \int_{0}^{\tau} \left[ X(i) - X_{\rm{ini}}(i)\right] 
\dot{M}(t) dt,
\label{yield}
\end{equation}

where $X(i)$ and $X_{\rm{ini}}(i)$ are the current and initial mass fractions of species $i$ respectively, $\tau$ is the stellar lifetime and $\dot{M}(t)$ is the mass-loss rate.
We provide yield tables formatted with the following columns; species ($i$), net yield $M_i$ in M$_\odot$, total mass expelled in winds $M^{\mathrm{wind}}(i)$ and production factor $\log_{10}(\langle X(i)\rangle/X_{\rm{ini}}(i)$),\footnote{As the initial composition of these models is scaled solar, our production factor $\log_{10}(\langle X(i)\rangle/X_{\mathrm{ini}}(i)$) is almost equivalent to [X/Fe].} where $\langle X(i)\rangle$ is the average mass fraction (the total amount of species $i$ expelled into the ISM divided by the total amount expelled) and $X_{\mathrm{ini}}(i)$ is the initial mass fraction. A sample format of these tables is shown in Table~\ref{tableappend2} in Appendix B.

To maximize utility and in order to allow comparison with other works our yields are presented in terms of production factors (Figures~\ref{productionfactor}, \ref{testcase} and \ref{compare}),  net yields (Figures~\ref{yields1} and \ref{yields2}), [X/Fe]\footnote{Where [A/B]=log$_{10}$($n$(A)/$n$(B))$_*$-log$_{10}$($n$(A)/$n$(B))$_\odot$} (Table~\ref{recapitulation}), total mass expelled (Figure~\ref{compare}) as well as weighted by the \cite{kro93} initial mass function (Sect.~\ref{sec-wei}).

Figure~\ref{productionfactor} shows the production factor of our standard models for Z=0.02, 0.008 and 0.004. We see large production of \chem{13}C, \chem{14}N, \chem{17}O and \chem{25}Mg, whilst \chem{15}N,  \chem{18}O and \chem{21}Ne are efficiently destroyed. 
When looking at specific isotopes which show large variation over this mass and metallicity range, the most noticeable are \chem{7}Li, \chem{24}Mg and \chem{25}Mg.
The model results show quite uniform features across the metallicity range considered, with no noticeable change in behaviour of the nucleosynthetic yields as they transition from massive AGB to super AGB stars.\footnote{The minimum mass for carbon ignition, $M_{\rm{up}}$, defines the transition from massive AGB to super AGB star. This quantity is metallicity dependent, with $M_{\rm{up}}$= 8.1 M$_\odot$, 7.7 M$_\odot$ and 7.3 M$_\odot$ for Z=0.02, 0.008 and 0.004 respectively, although this depends critically on how one treats the border of convective cores.} This regularity in production is due to HBB dominating 3DU and all other nucleosynthetic processes in these stars.
 
As mentioned in the introduction, a major aim of this study was to find a clear observational signature of a super AGB star compared to their slightly lower mass counterparts. From Figure~\ref{productionfactor}, we see no obvious suitable isotopic candidate, however, with the significant increase in our s-process proxy particle $g$, these heavier than iron isotopes may yet hold a key to identification of super AGB stars, with discussion of $g$ in Section~\ref{sprocess}.

Next we discuss the yields of specific species, and then investigate the model uncertainties due to mass-loss rates and variations in the mixing length parameter $\alpha_{\rm{mlt}}$. We then examine all presented yields in a broader context by using IMF weighted yields and comparing these to lower mass AGB models. The uncertainties related to the possible use of extrapolated thermal pulses beyond convergences issues at the tip of the (S)AGB phase are then discussed. This section then ends with comparison to other results in the literature.

\subsection{Standard Yields}

\subsubsection{Light Elements He, Li}\label{subsubhheli}

There is a substantial increase in \chem{4}He, primarily due to the effects of 2DU but supplemented by efficient HBB and 3DU.
We find mass fraction values in excess of 0.350 and $\approx$ 0.5 M$_\odot$ of \chem{4}He is produced in every super or massive AGB star.

Lithium is depleted during both 1DU and 2DU and produced in significant amounts at the start of the super AGB phase, via the \cite{cam71} mechanism. Values of up to log$\epsilon$(\chem{7}Li) $\sim$ 4.4 are reached, with the maximum values increasing with the initial mass and decreasing with the metallicity.
It has been shown by a variety of authors \cite[e.g.][]{sac92a,tra01,ven10a} that lithium yields are strongly dependent on both the timing and rate of mass-loss.
If the mass-loss rate is sufficiently rapid at the beginning of the (S)AGB phase, before significant \chem{7}Li destruction takes place, the net yield will be high. For all metallicities, the yield increases with initial mass from negative to positive values.  
We see overall positive net yields for the Z=0.02 models due to a more rapid average mass-loss rate as well as the more moderate $T_{\rm{BCE}}$.  

\subsubsection{C, N, O and F}

Although \chem{12}C increases at the surface due to 3DU it is subsequently depleted via HBB. 
Even with efficient third dredge-up, \chem{12}C yields are negative for all Z=0.02 and Z=0.008 models, whilst only the lowest mass Z=0.004 model shows a slight positive yield as a consequence of the larger comparative effect of the later 3DU events when HBB has ceased.  
The less abundant isotope of carbon, \chem{13}C, is largely produced via CN cycling of \chem{12}C; its yields increase as a function of the metallicity and are positive for all models here. 

The surface \chem{14}N abundance is enhanced through first and second dredge-up. It is also produced in large quantities via HBB of the 3DU product \chem{12}C.
\chem{15}N is destroyed very efficiently at the base of the convective envelope which leads to quite large negative yields. In Figure~\ref{productionfactor} the production factor of \chem{15}N becomes less negative with increasing initial mass $M_{\rm{ini}}$ and decreasing metallicity as its equilibrium value with \chem{14}N is larger at the higher temperatures reached.

\chem{16}O is depleted during both 1DU and 2DU and further consumed via ON cycling during HBB. There is some replenishment due to 3DU but this is not enough to lead to a net positive yield. 
\chem{17}O presents a very large increase due to CNO cycling in both the H shell and at the base of the convective envelope.
The surface abundance of \chem{18}O is depleted during 1DU, and also in standard 2DU events. The more massive super AGB models undergo corrosive 2DU \cite[e.g.][]{gil13} whereby the convective envelope penetrates into the top of the thick helium shell and dredges up the \chem{18}O which had been produced via $\alpha$ captures on \chem{14}N. Even in models with this large enrichment, the efficient HBB destroys \chem{18}O which always results in negative yields.

The production of \chem{19}F involves a multi-step process which relies on both 3DU and thermal pulse nucleosynthesis \cite[for details see][]{mow96,lug04}. This leads to surface enrichment from 3DU, however it is easily destroyed by HBB at high temperatures and the yield is therefore always strongly negative in these super/massive AGB stars.

 \subsubsection{Ne-Na, Mg-Al, Si, P and S}

The net yields of \chem{20}Ne are positive (albeit very slightly) for Z=0.004 and 0.008 in the entire mass range, whilst only the two most massive models for Z=0.02 show positive yields. This very slight production is via 3DU.   
 The yields of \chem{21}Ne are always negative as it is rapidly burnt by HBB although 3DU can enrich the surface near the end of the evolution when HBB has ceased.
Production of \chem{22}Ne in a thermal pulse is via \chem{14}N($\alpha,\gamma$)\chem{18}F($\beta^+,\nu$)\chem{18}O($\alpha,\gamma$)\chem{22}Ne, and the \chem{19}F($\alpha$,p)\chem{22}Ne channel. This is subsequently dredged-up to the surface and slightly depleted thereafter by HBB. 

 The \chem{23}Na yields are positive in all models, due almost entirely to the large increase from 2DU. 
During HBB \chem{23}Na is produced from proton captures on \chem{22}Ne, but for the considered  $T_{\mathrm{BCE}}$, the destruction channel dominates. The yield is lower at lower metallicities due to the more efficient destruction from hotter HBB. 

The three magnesium isotopes in intermediate-mass AGB stars have been extensively investigated \cite[e.g.][]{arn99,kar03,den03,kar06a,sie10,ven11b} mainly to explain globular cluster anomalies however the isotopic abundances of Mg also have relevance to Lyman alpha systems \citep{aga11}.  

Both HBB and 3DU events change the surface abundance and yields of magnesium isotopes in super and massive AGB stars. Firstly in the convective pulse, the \chem{22}Ne ashes undergo an $\alpha$ capture to form either \chem{25,26}Mg while at the same time, a fraction of this newly created \chem{25}Mg can capture a free neutron to form \chem{26}Mg. So in effect, the intershell region after a thermal pulse is very enriched in \chem{26}Mg and to a lesser extent in \chem{25}Mg.
After \chem{25,26}Mg have been dredged up to the surface they suffer additional nuclear burning due to HBB.
As $T_{\mathrm{BCE}}$ exceeds $\approx$ 90 MK there is a large production of \chem{25}Mg at the expense of \chem{24}Mg. The net yield of \chem{25}Mg is governed by its production through HBB from \chem{24}Mg (Fig.~\ref{yields2}). The \chem{26}Mg net yield in this figure shares a striking similarity with both $g$ and \chem{60}Fe. The lockstep changes in these isotopes imply that they have the same origin, suggesting that \chem{26}Mg is almost purely a product of 3DU in this mass and metallicity range.

\chem{26}Al is created in the envelope via proton capture reactions on \chem{25}Mg and destroyed via a further proton capture leading to \chem{27}Si if $T_{\mathrm{BCE}}> $ 100 MK. We discuss \chem{26}Al in more detail in Section~\ref{radio}.

In the considered metallicity range there is only a slight \chem{27}Al production, primarily from the dredge-up of material created via \chem{23}Na($\alpha,\gamma$)\chem{27}Al and \chem{26}Mg(n,$\gamma$)\chem{27}Mg($\beta^+,\nu$)\chem{27}Al. The contribution from HBB to this isotope is minimal.
The production factor for silicon, sulfur and phosphorous isotopes for the standard models presented here is negligible, in the order of 10$^{-2}$ dex.

\subsubsection{S-process tracer: $g$}\label{sprocess}

Whilst the detailed nucleosynthesis of s-process production is outside the scope of this work (to be explored in Lau et al. (2013, in preparation)), our fictitious $g$, the s-process proxy isotope, gives an estimate of the amount of heavier than Fe material produced. Its yield is reliant on many factors such as neutron flux and source (in this case primarily \chem{22}Ne($\alpha$,n)\chem{25}Mg), availability of seed nuclei, overlap factor of successive convective intershells as well as the 3DU efficiency and mass loss rate that participate to the release of this element in the ISM.
In Figure~\ref{productionfactor} a large enhancement (up to 1.2 dex) is seen in $g$, with generally the lower metallicity and less massive models having the largest production factor.

\begin{figure*}
\resizebox{\hsize}{!}{\includegraphics{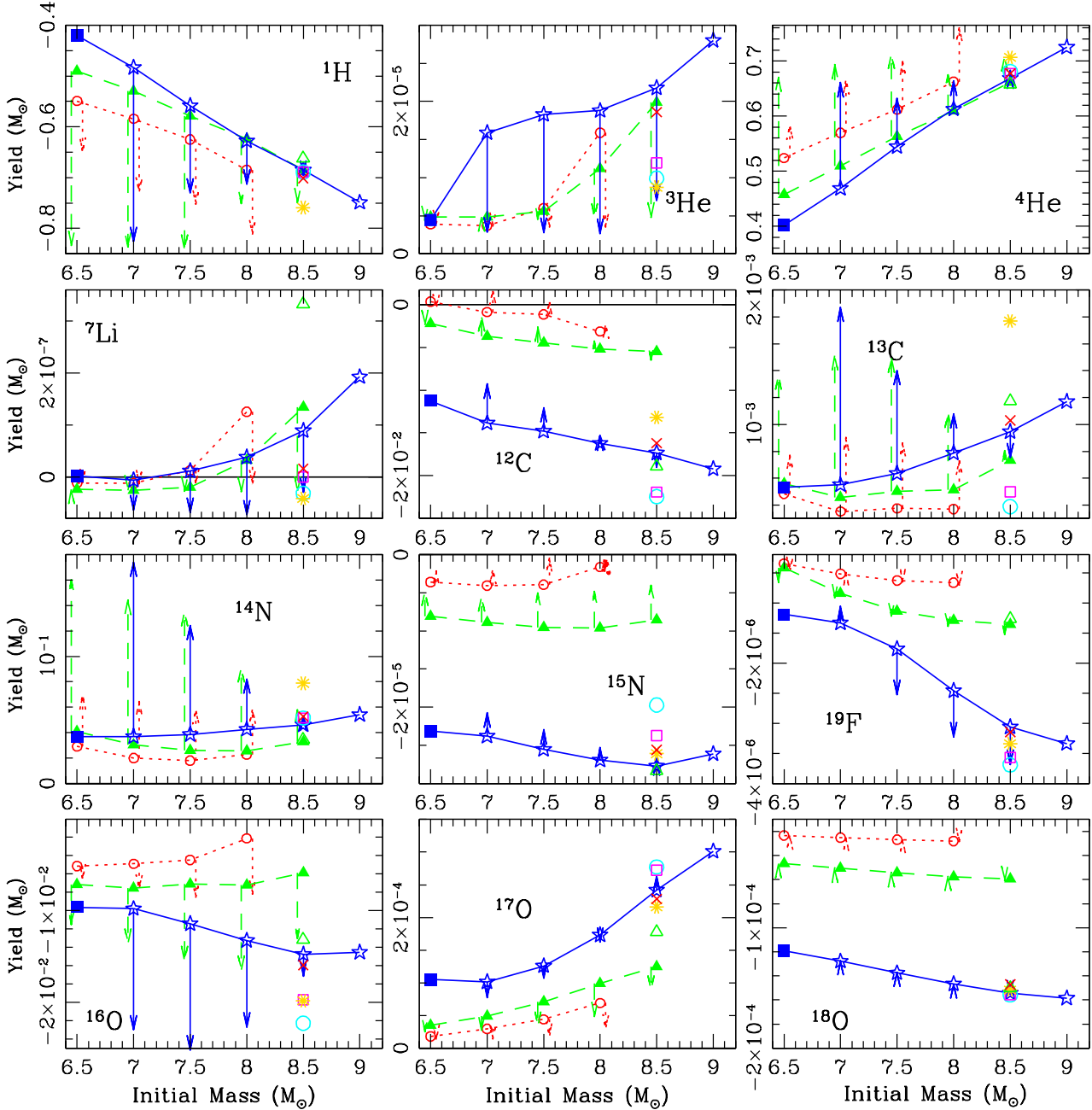}}
\caption{Net yield (M$_\odot$) as a function of initial mass for metallicities Z=0.02, 0.004 and 0.008. Calculations for Z=0.02 are shown with open stars and solid line, Z=0.008 with filled triangles and dashed line whilst Z=0.004 models are represented by open circles with dotted line. The test cases for 8.5 M$_\odot$ Z=0.02 are also shown, with symbols as per Figure~\ref{testcase}. The arrows represent the change due to the use of the delayed super-wind phase for the mass-loss rate. Bold arrows represent the length of the arrow $\times$ 10. Gold diamonds represent R75 yields divided by 5. The filled blue square represents the Z=0.02 6.5 M$_\odot$ model result from Karakas (2010).}
\label{yields1}
\end{figure*} 

\begin{figure*}
  \resizebox{\hsize}{!}{\includegraphics{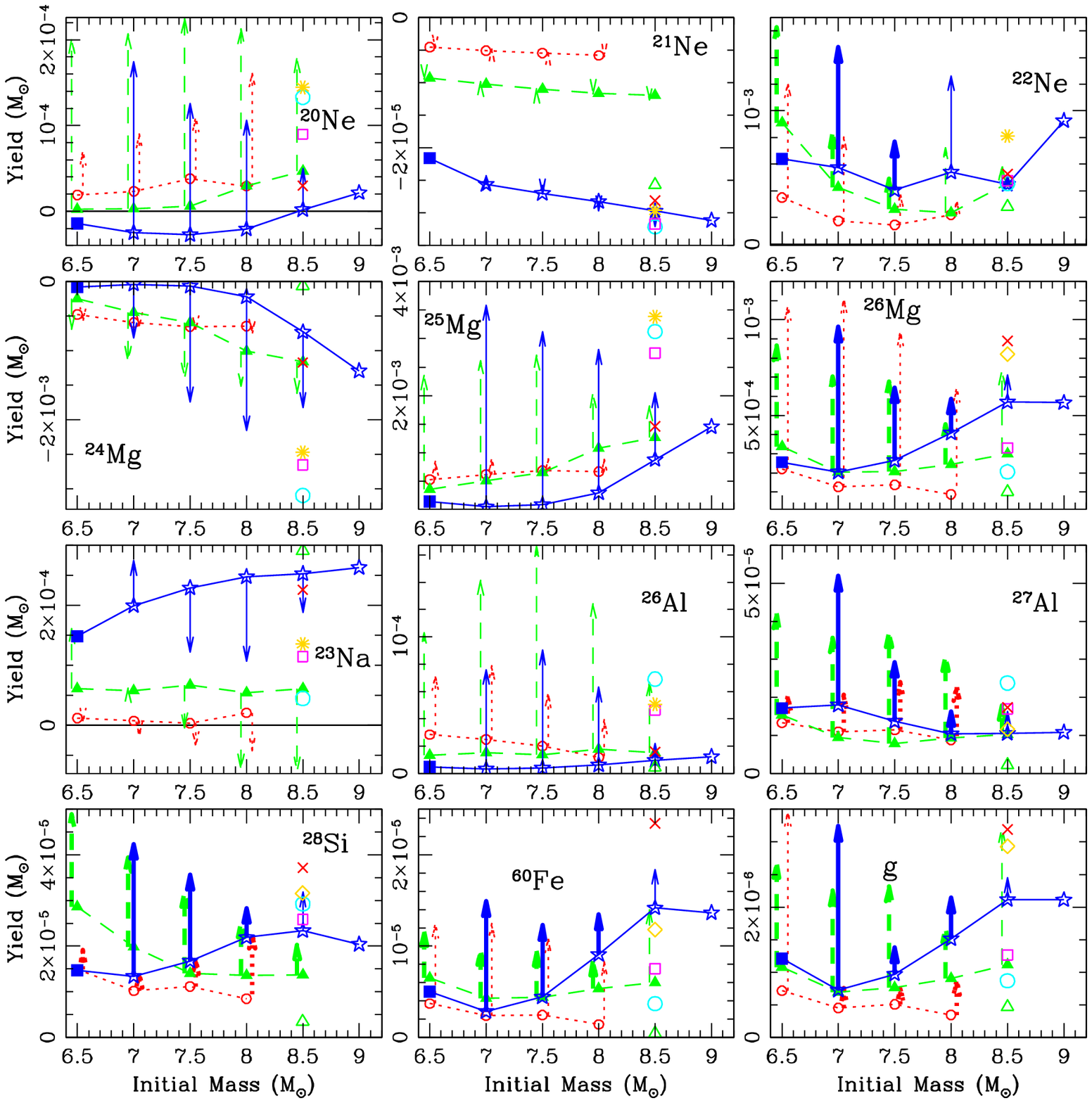}}
 \caption{Net yield (M$_\odot$) as a function of initial mass for metallicities Z=0.02, 0.004 and 0.008. Symbols as per Figure~\ref{yields1}.}
 \label{yields2}
 \end{figure*}

\subsection{Effect of Mass-loss rate}\label{sec-ml}
\begin{figure}
 \resizebox{\hsize}{!}{\includegraphics{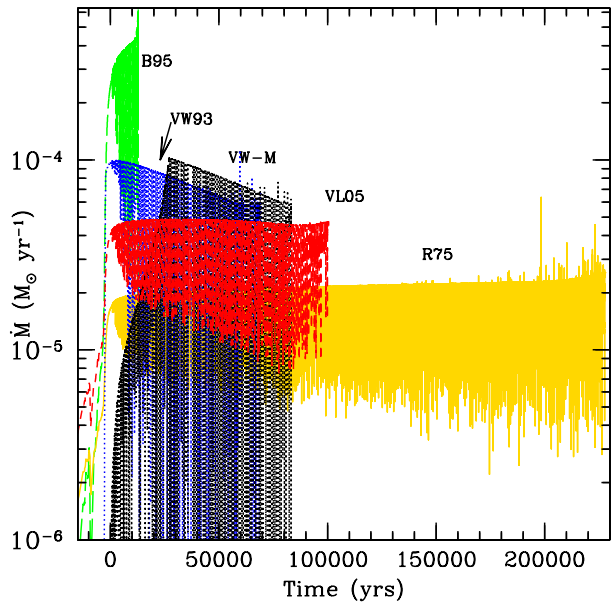}}
 \caption{Mass-loss rate in M$_\odot$ yr$^{-1}$ as a function of time during the thermally pulsing super AGB phase for the set of 8.5 M$_\odot$ Z=0.02 models.}
 \label{masslossrate}
 \end{figure}

 We have investigated the effect of differing mass-loss rates on the evolution and nucleosynthesis by running a set of test models for a 8.5 M$_\odot$ Z=0.02 star using the four most common mass-loss formulae for massive AGB stars (Fig.\ref{masslossrate}). In order of decreasing mass-loss rate these are: \cite{blo95} using $\eta$=0.1, denoted B95, \cite{van05} denoted VL05, \cite{vas93} denoted VW93 (our standard case) and \cite{rei75} with $\eta$=5, denoted R75. 
 There are two main ways in which the mass-loss rate affects the stellar yields. Firstly it simply alters the duration of the thermally pulsing phase resulting in less/more thermal pulses and their associated third dredge-up events. Secondly it changes the temperature at the base of the convective envelope, with a more rapid mass-loss rate leading to a lower temperature, resulting in a change to the HBB nucleosynthesis.

 In Table~\ref{testcasetable} we include information such as the number of thermal pulses and the amount of dredged-up material for the mass-loss test cases which reflect the effects mentioned above. The variation in the number of thermal pulses and duration of the (S)-AGB is quite remarkable. The number of TPs range from 29 (B95) to 381(R75), whilst the super AGB lifetime is $\sim$ 20 times longer when the R75 rate is used compared to B95.

\begin{figure*}
 \resizebox{\hsize}{!}{\includegraphics{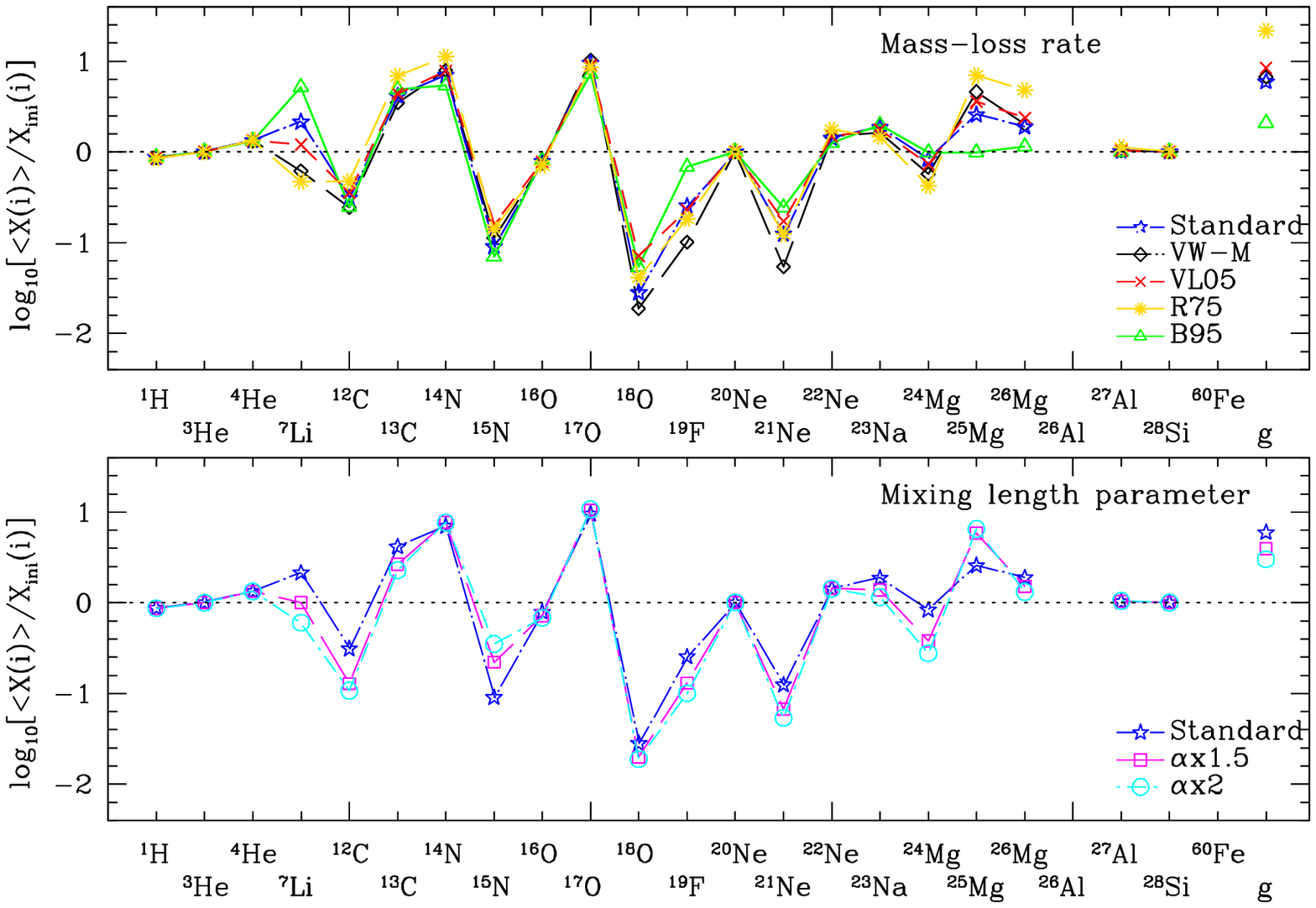}}
 \caption{Production factors for the 8.5 M$_\odot$ Z=0.02 stellar model test cases with differing mass-loss rate (top panel) and mixing length parameter $\alpha_{\rm{mlt}}$ (bottom panel).}
 \label{testcase}
 \end{figure*}

The B95 rate is used quite frequently in AGB calculations \cite[e.g.][]{her04b} but due to its very large mass-loss rates for high luminosity objects, it has been more recently used with a lower value of $\eta$ such as $\eta$=5$\times$10$^{-4}$ \cite[]{her12}, or $\eta$= 0.02 \cite[]{ven11b}, chosen to match the proportion of lithium rich giants in the LMC \citep{ven00}(cf. $\eta$ =0.1 in our case).
The R75 mass-loss rate, which is most commonly used for the RGB, was originally shown to be a reasonable choice of mass-loss for the AGB with $\eta$=5 \citep{gro93}. However its accuracy for AGB stars is now in doubt \cite[e.g.][]{gro09}.
 As the VL05 and VW93 rates are based on observations of either luminous AGB stars or super-giants they would seem the most appropriate for massive AGB stars. The results for these two rates are reasonably comparable in their global characteristics, with $\sim$ 30 per cent difference in number of TPs and $\tau_{(S)AGB}$.
The top panel of Figure~\ref{testcase} shows the production factor for our selected wind prescriptions. Models with slower mass-loss rates have substantially more 3DU episodes which further enrich the surface abundances of \chem{19}F, \chem{25}Mg, \chem{26}Mg, g and to a lesser extent \chem{13}C, \chem{14}N and \chem{24}Mg. 
\chem{7}Li is also shown to be highly dependent on the mass-loss rate as discussed in Section~\ref{subsubhheli}, with the more rapid mass-loss in the early portion of the AGB phase leading to larger yields.
In the B95 case, the yield of \chem{19}F is only slightly negative, by virtue of rapid evolution and expulsion of the envelope before \chem{19}F has been destroyed via HBB.
Overall, for our selected model, whilst the two most extreme mass-loss rates produce variations by up to a factor of 10 in some isotopes e.g. \chem{25}Mg, \chem{26}Mg, the VW93 and VL05 yield values vary considerably less, by at most a factor of two. The net yield of selected isotopes for all of these mass-loss rates can be found in Figures~\ref{yields1} and \ref{yields2} using the same symbols as Figure~\ref{testcase}.

Our second test for the mass-loss comprises computations of a set of models with the VW93 standard mass-loss rate modified to include the delayed onset of the super-wind phase. To account for the O-rich optically visible long period variable stars with pulsation periods\footnote{The pulsation period (in the fundamental mode) is a function of stellar mass and radius with log P (days) = -2.07 + 1.94 log R/R$_\odot$ - 0.9 log M/M$_\odot$.} up to 750 days, \cite{vas93} suggested a delay in the mass-loss to the super-wind phase for massive AGB stars (their equation 5\footnote{log $\dot{M}$ = -11.4 + 0.0125 [P $\rm{(days)}$ - 100($\rm{M}$/M$_\odot$ -2.5)]}). This equation has an initial mass dependent delay in addition to the 500 day delay in the standard formula. Using the mass-loss rate in this form would result, for the most massive super AGB stars, in super-wind delays until periods in excess of 1350 days! This is a result not borne out by observations. We have therefore used a modified version of the VW93 rate (hereafter VW-M\footnote{Our modified formula (VW-M) reduces to 350 the [100($\rm{M}$/M$_\odot$ -2.5)] term in VW93 equation 5.}) to take into account the work of \cite{deb10} who find a levelling off or saturation of the mass-loss rate at $\sim$ 850 day period. Hence we use this 850 day period limit for all masses considered here, similar to the approach of \cite{kar12} who instead use 700 or 800 day delays. 

We have also introduced a secondary criterion for the onset of the super-wind phase, this being when the pulsation period is in excess of 500 days and the surface abundance ratio C/O exceeds unity. The C/O criterion was added to account for the lack of variable compositional low temperature molecular opacities in this study. Unlike our standard models, in which the inclusion of these opacities are expected to only have a small/possibly negligible affect, for these slower mass-loss rate models, especially at lower metallicity, the surface composition becomes carbon rich comparatively early on the TP-(S)AGB phase, in some cases prior to the super-wind phase. The models affected by this secondary criterion are the 6.5, 7 and 7.5 M$_\odot$ models of metallicities Z=0.008 and Z=0.004. The C/O ratio exceeds unity when the period exceeds (in order of increasing initial mass) $\sim$ 700, 750 and 820 days for the Z=0.008 models and $\sim$ 650, 700 and 800 days in the Z=0.004 models.
With inclusion of variable compositional molecular opacities, the \textit{maximum} stellar TP-(S)AGB lifetimes are strongly dependent on the envelope C/O ratio, and even with efficient HBB eventually, given a slow enough mass loss rate we expect all models to become carbon rich. This limits the amount of intershell material, in particular s-process enriched material that can be dredged to the surface in super and massive AGB stars.
 
As the mass increases, the results for standard versus modified VW93 prescriptions converge, with the most massive models already achieving a pulsation period in excess of 850 days, hence entering the super-wind stage at the beginning of the thermally pulsing AGB phase.  
The maximum effect of using the delayed super-wind would be for the less massive AGB stars in the mass range $\approx$ 6.5-7 M$_\odot$.    

\begin{figure}
\resizebox{\hsize}{!}{\includegraphics{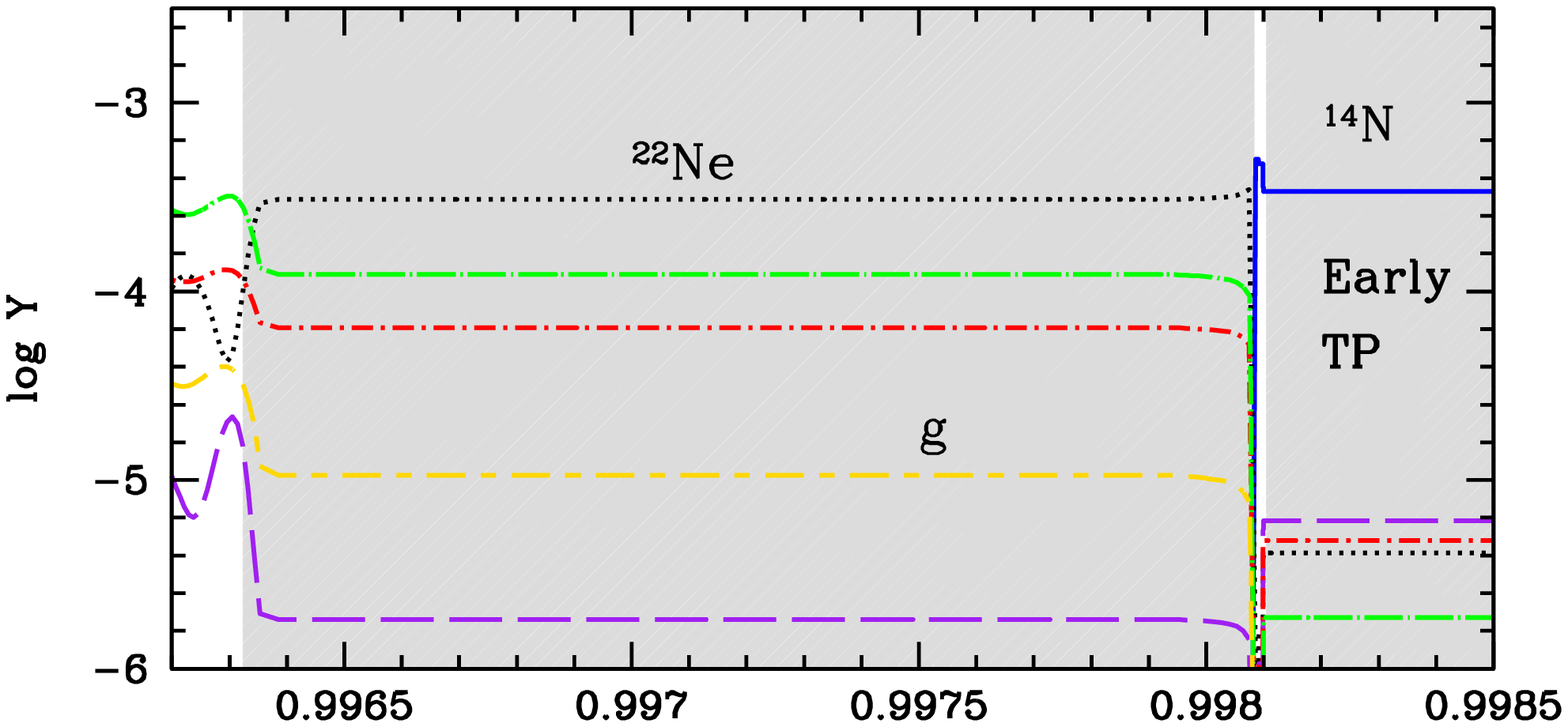}}
\resizebox{\hsize}{!}{\includegraphics{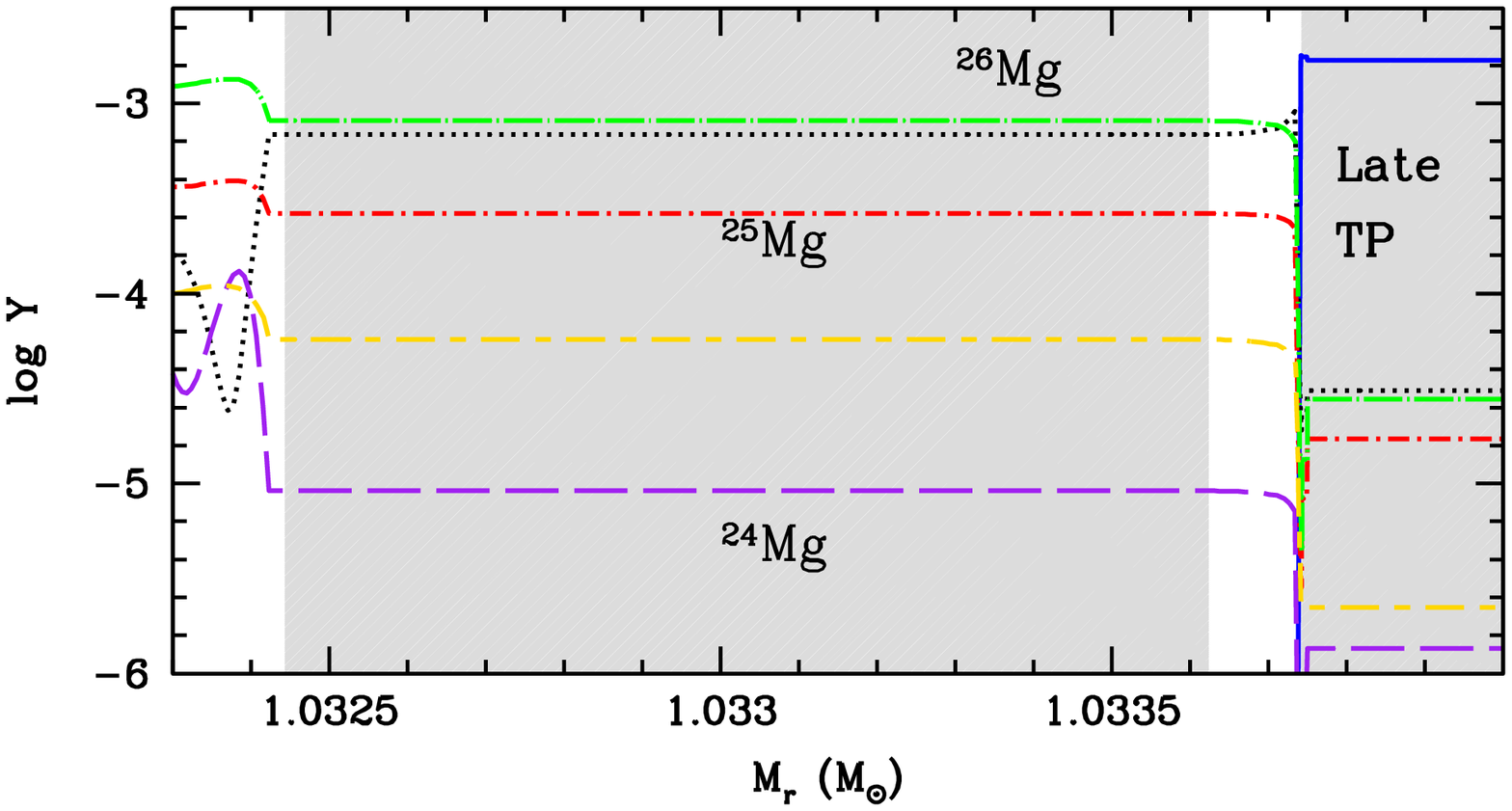}}
 \caption{Composition profiles for the 7 M$_\odot$ Z=0.008 VW-M model during the
 30th (top panel) and 300th (bottom panel) thermal pulses.  Grey shading represents convective regions.}\label{fig-inter} \end{figure}

\begin{figure*}
\begin{center}
\resizebox{\hsize}{!}{\includegraphics{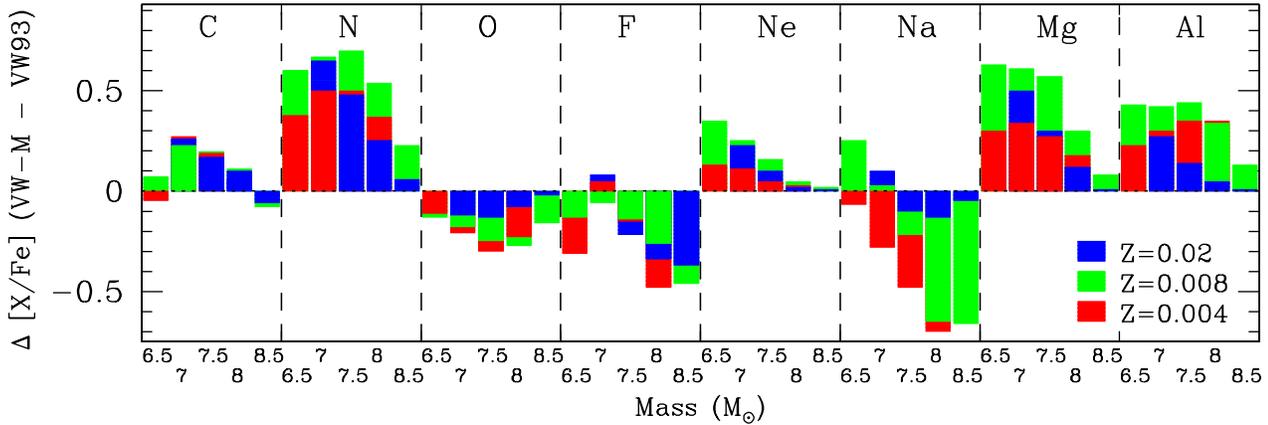}}
\caption{Elemental abundance differences between VM-M model yields and our standard (VW93) model yields in [X/Fe]. The masses 6.5 M$_\odot$ and 8.5 M$_\odot$ only have yield results for two metallicities.}
\label{xfe}\end{center}
\end{figure*}

The results for these models can be found in the online yield tables as well as being plotted as arrows in Figures \ref{yields1}, \ref{yields2} and \ref{fig-giants}. 
The use of the VM-M prescription increases the number of thermal pulses as well as the duration of the TP-(S)AGB phase by up to a factor of 10 for the lower mass models as seen in Table~\ref{tpsagb}. 

Clearly seen in Figures~\ref{yields1} and \ref{yields2} is the significant increase in the yields of \chem{4}He, \chem{14}N, \chem{13}C and the heavy magnesium isotopes \chem{25}Mg and \chem{26}Mg with decreasing mass-loss rate.
There is also a large production of the radioactive isotope \chem{26}Al. 
The very large increase in the \chem{25}Mg and \chem{26}Mg is a direct consequence of the higher \chem{14}N in the envelope, which is then transmuted to \chem{22}Ne, \chem{25}Mg and \chem{26}Mg during the TP. The helium burning intershell convective zone temperature also increases quite
substantially due to the much longer thermally pulsing phase and larger core mass. These effects together result in up to a factor of 5 increase in the abundance of the heavy magnesium isotopes generated within late TPs compared to the early TPs. As seen in Figure~\ref{fig-inter} which shows the intershell abundances of the 30th TP and 300th TP for the 7 M$_\odot$ Z=0.008 VW-M. Combining these high intershell abundances with the larger number of TPs explains the very high Mg production found in these models.

Other isotopes noticeably increased when the VW-M rate is used are \chem{27}Al and \chem{28}Si. In Figure~\ref{xfe} we see the variation in elemental yields [X/Fe] between the VM-M and VW93 model yields. The [X/Fe] for these models can be found in Table~\ref{recapitulation}.

The large increase in some of the isotopic yields such as \chem{14}N, \chem{25}Mg and \chem{26}Mg may possibly rule out such a slow mass-loss rate for these massive AGB stars. This would be interesting to test in a galactic chemical evolution model.
The longer lifespan of the thermally pulsing phase for these very luminous AGB stars may be a possible observational constraint.

\subsection{Effect of varying the Mixing length theory parameter $\alpha_{\rm{mlt}}$}

 Our second set of test models explore the effects of changing the mixing length parameter $\alpha_{\rm{mlt}}$ during the thermally pulsing super AGB phase. We have again used our representative super AGB model (8.5 M$_\odot$ Z=0.02 with VW93 mass-loss) with a standard value of $\alpha_{\rm{mlt}}$ and then models with this value multiplied by 1.5 and 2. These quite drastic changes in $\alpha_{\rm{mlt}}$ have been motivated by the inherent uncertainties in the MLT values, and lack of evidence for the appropriateness of using the solar calibrated values for other phases of evolution or metallicity \cite[e.g.][]{sac91,lyd93,chi95}. There is also a growing amount of computational/theoretical \cite[e.g.][]{ven05,woo11a} and observational \citep{mcs07} work which suggest larger values than traditionally used are required during the AGB phase.

\begin{figure}
 \resizebox{\hsize}{!}{\includegraphics{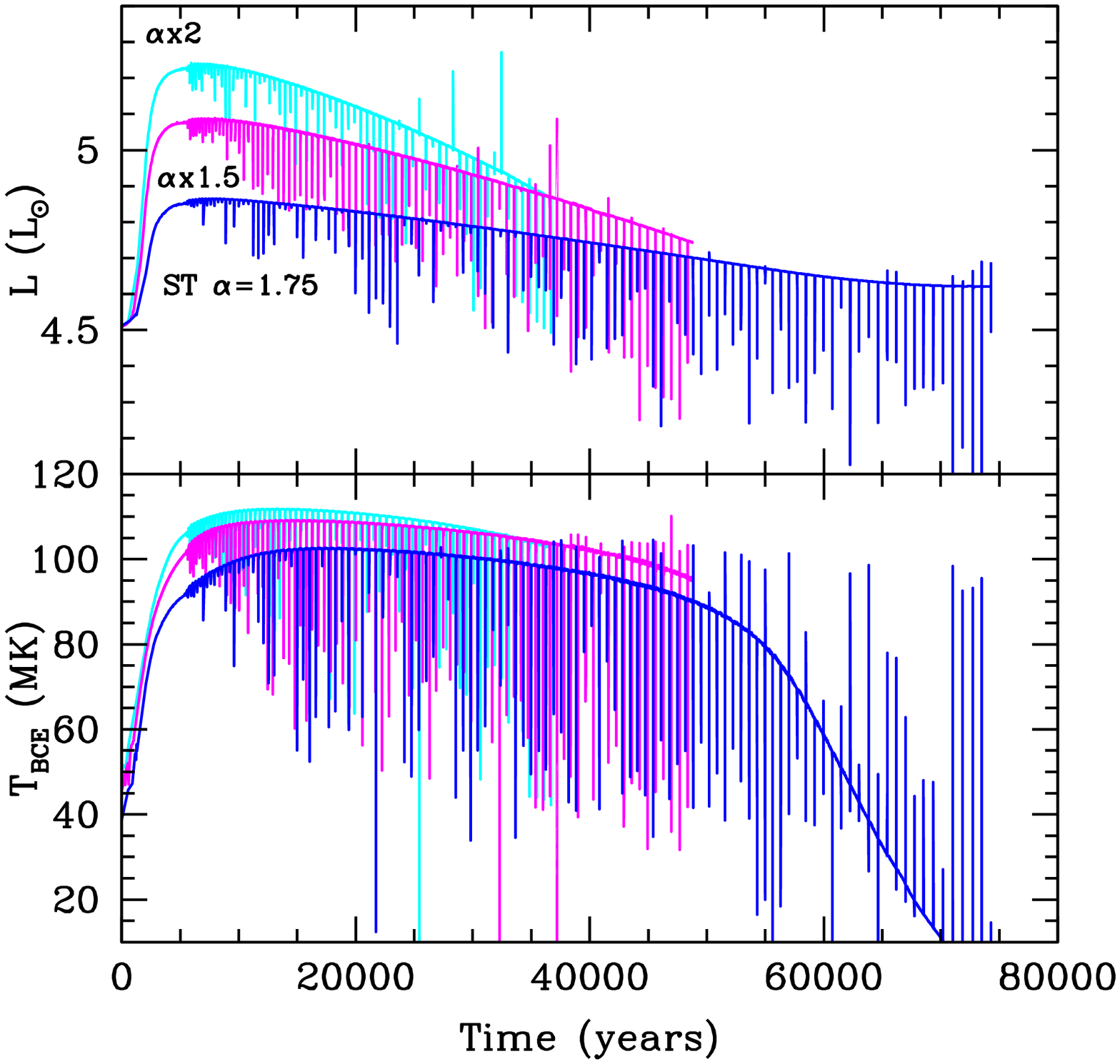}}
 \caption{Total luminosity (top panel) and temperature at the base of the
   convective envelope (bottom panel) as a function of time during the
   thermally pulsing super AGB phase for the set of $\alpha_{\mathrm{mlt}}$ tests for the 8.5 M$_\odot$ Z=0.02 models.}
 \label{alpha}
 \end{figure}

As shown in Figure~\ref{alpha}, with increasing $\alpha_{\rm{mlt}}$ there is an increase in $T_{\rm{BCE}}$ and luminosity L. This leads to a subsequent increase in the mass-loss rate, shorter lifetime of the (S)AGB phase and fewer 3DU events as seen in Table~\ref{testcasetable}. Another consequence of a larger value of $\alpha_{\rm{mlt}}$ is that the stellar model proceeds to a smaller envelope mass prior to convergence issues halting their calculations (for more details refer to \cite{lau12}).
We note that unlike in the super AGB calculation from \cite{her12} we do not find an increase in the efficiency of 3DU with increasing $\alpha_{\rm{mlt}}$. 

A surprising result from Fig.~\ref{testcase} is the \chem{7}Li behaviour. Whilst one may expect that a greater mass-loss rate would lead to a larger lithium production this is not the case because as $\alpha_{\rm{mlt}}$ increases, \chem{7}Li is also depleted more effectively as a result of a larger $T_{\rm{BCE}}$.  
Due to the more efficient mixing as well as higher $T_{\mathrm{BCE}}$, \chem{19}F is further depleted and \chem{25}Mg production enhanced at the expense of \chem{24}Mg. Quantitatively, for \chem{25,26}Mg, we see at most a factor of 3 change due to variation of the MLT parameter. Note however that the effect of increasing $\alpha_{\rm{mlt}}$ would be largest in the 6-7 M$_\odot$ mass range where $T_{\mathrm{BCE}}$ reaches the threshold of activation of the Mg-Al cycle.
 Another key difference noted with the increase in $\alpha_{\rm{mlt}}$ is the large increase in \chem{26}Al as seen in Figure~\ref{al26} and further discussed in Section~\ref{radio}. Although increasing $\alpha_{\rm{mlt}}$ leads to more efficient ON cycling, the C/O ratio still stays well below unity for the whole TP-(S)AGB phase.
 
\begin{table} \begin{center}
\caption{Selected model characteristics for the 8.5 M$_\odot$ Z=0.02 models exploring the effects of varying of mass loss rate and MLT parameter. Variables as in Table~\ref{tpsagb}. See text for details.}
\label{testcasetable}\setlength{\tabcolsep}{1.5pt} 
 \begin{tabular}{lcccccrc}\hline \hline
&$T_{\rm{BCE}}^{\rm{Max}}$&$M_{\rm{Dredge}}^{\rm{Tot}}$ & $M_{\rm{2DU}}$& $M_{\rm{C}}^{\rm{F}}$ &$M_{\rm{env}}^{\rm{F}}$ &  $N_{\rm{TP}}$  & $\tau_{\rm{(S)AGB}}$ \\
&(MK)&(M$_\odot$)&(M$_\odot$)&(M$_\odot$)&(M$_\odot$)&&(yrs)\\\hline
 \multicolumn{8}{c}{Z=0.02 8.5 M$_\odot$} \\ \hline
VW93                       &106 &3.64(-2) & 1.14 &1.15 &1.78 &110[35]  & 6.60(4) \\
VW-M                       &106 &4.75(-2) & 1.14 &1.16 &1.87 &149[40]  & 8.63(5) \\
VL05                       &107 &5.66(-2) & 1.14 &1.16 &1.93 &165[54]  & 9.87(4) \\ 
B95                        &98  &4.34(-3) & 1.14 &1.15 &1.56 &29[2]    & 1.16(4) \\ 
R75                        &107 &1.35(-1) & 1.14 &1.17 &2.01 &381[174] & 2.26(5) \\
$\alpha_{\rm{mlt}}\times1.5$&109 &2.40(-2) & 1.14 &1.15 &1.36 &84[28]   & 4.38(4) \\ 
$\alpha_{\rm{mlt}}\times2$  &112 &1.62(-2) & 1.14 &1.15 &1.17 &65[18]   & 3.06(4) \\
\hline
\end{tabular}
 \end{center} \end{table}

\subsection{Radioactive isotopes - \chem{26}Al, \chem{60}Fe}\label{radio}

The radioactive isotopes \chem{26}Al and \chem{60}Fe\footnote{\chem{26}Al has a terrestrial half life $\tau_{1/2}$ of 0.717 Myrs whilst \chem{60}Fe is more long lived with $\tau_{1/2}$ $\approx$ 2.2 Myrs.} are of great interest because their $\gamma$-ray emission can be observed over very large distances \citep{pra96} and give an instantaneous snapshot of the on-going nucleosynthesis. 

These two isotopes are also some of the most studied of the short lived radioactive (SLR) nuclei in the early solar system. We have recently suggested, using the models presented here, that super AGB stars may have been the source of these SLR isotopes \citep{lug12b}. 
Both radioactive isotopes come from different regions within an AGB star. The \chem{26}Al is a HBB product, whilst \chem{60}Fe is created within the thermal pulse via neutron captures and later dredged-up to the surface.

\begin{figure}
\resizebox{\hsize}{!}{\includegraphics{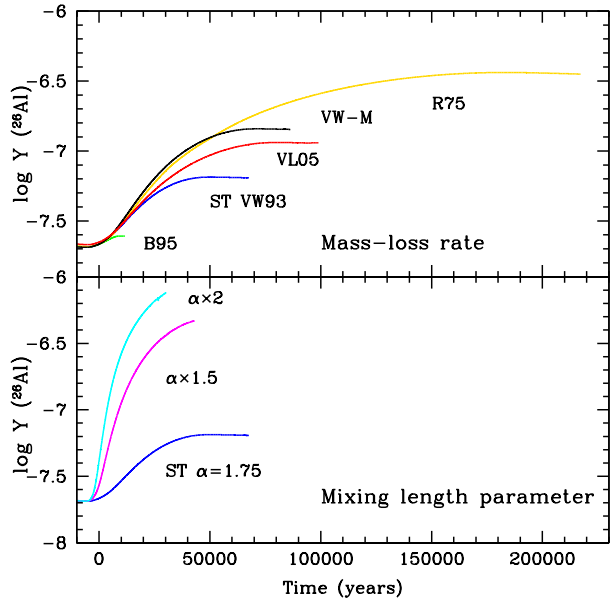}}
\caption{\chem{26}Al surface abundance in log mole fraction Y as a function of time for the series of  8.5 M$_\odot$ Z=0.02 stellar model test cases.}
\label{al26}
\end{figure}

The surface abundance of \chem{26}Al\footnote{Although here we discuss the
  yields of \chem{26}Al it should be remembered that this \chem{26}Al will
  subsequently decay to \chem{26}Mg. In our online tables we include these
  species separately, but the net yield contribution from \chem{26}Al
  should be added to \chem{26}Mg.} for the variety of test cases for the
8.5 M$_\odot$ Z=0.02 can be seen in Figure~\ref{al26}. From the top panel
we see there is a very strong dependence of the production of \chem{26}Al
on the duration of the thermally pulsing phase which is indirectly related to our choice of $\dot M$. This result may thus potentially help constrain the mass-loss rate.
The bottom panel clearly shows the significant change in production of \chem{26}Al with increased $\alpha_{\rm{mlt}}$ (and therefore higher $T_{\mathrm{BCE}}$).
Whilst 3DU does not directly increase \chem{26}Al being an efficient neutron poison it is strongly depleted during a thermal pulse) the dredge-up of \chem{25}Mg then burnt via HBB can lead to a larger production of \chem{26}Al.

The present-day equilibrium galactic value of \chem{26}Al is estimated to be 2.8 M$_\odot\pm$ 0.8 M$_\odot$ \citep{die06a}. Although the major component of galactic \chem{26}Al (as well as \chem{60}Fe) is thought to come from massive stars \citep{lim06}, there are still large uncertainties due to reaction rates \cite[e.g.][]{tur10,ili11}.

The overall contribution of low and intermediate-mass AGB stars to the galactic inventory of \chem{26}Al has been discussed by a variety of authors \cite[e.g.][]{pra96,mow00,lug08} whilst \cite{sie08b} (hereafter S08) were the first to include the contribution from super AGB stars. They found that super (and massive AGB) stars contribute at most $\approx$ 0.3 M$_\odot$, which represents about 10 per cent of the galactic value.

Using our standard VW93 models we calculate (following the method outlined in \cite{pra96}) the overall contribution from solar metallicity super AGB stars (masses $\sim$ 8-9 M$_\odot$) to be $\sim$ 0.05 M$_\odot$.
If we include the total contribution from super AGB stars \textit{and} AGB stars for the mass range 3-9 M$_\odot$ (using stellar AGB yields from \cite{kar10} for masses $\leq$ 6.5 M$_\odot$) the values increases slightly to $\sim$ 0.12 M$_\odot$. 
If either the delayed super-wind models, or the increased $\alpha_{\mathrm{mlt}}$ model, are considered they do increase the galactic \chem{26}Al contribution in agreement with the 0.3 M$_\odot$ inferred by S08.

The large uncertainties in the reaction rates involved in \chem{26}Al production/destruction result in a quite large variation in \chem{26}Al yields for massive AGB stars \cite[e.g.][]{izz07,van08} however S08 found these reaction rate uncertainties had less of an affect on \chem{26}Al yields for super AGB stars. 
As in S08 we also find that the large uncertainty in convective mixing
efficiency (and especially how this translates to variations in
$T_{\rm{BCE}}$) is a major uncertainty in super AGB \chem{26}Al
yields. However, our delayed super-wind nucleosynthetic grid also
highlights and confirms the importance of the mass-loss rate to the \chem{26}Al yields as stressed by \cite{mow00}.
In Figure~\ref{yields2} the net yield of \chem{26}Al and \chem{60}Fe for all models is shown. 

\subsection{Weighted Yields}\label{sec-wei}

\begin{figure*}
\begin{center}
 \includegraphics[width=8.5cm,angle=0]{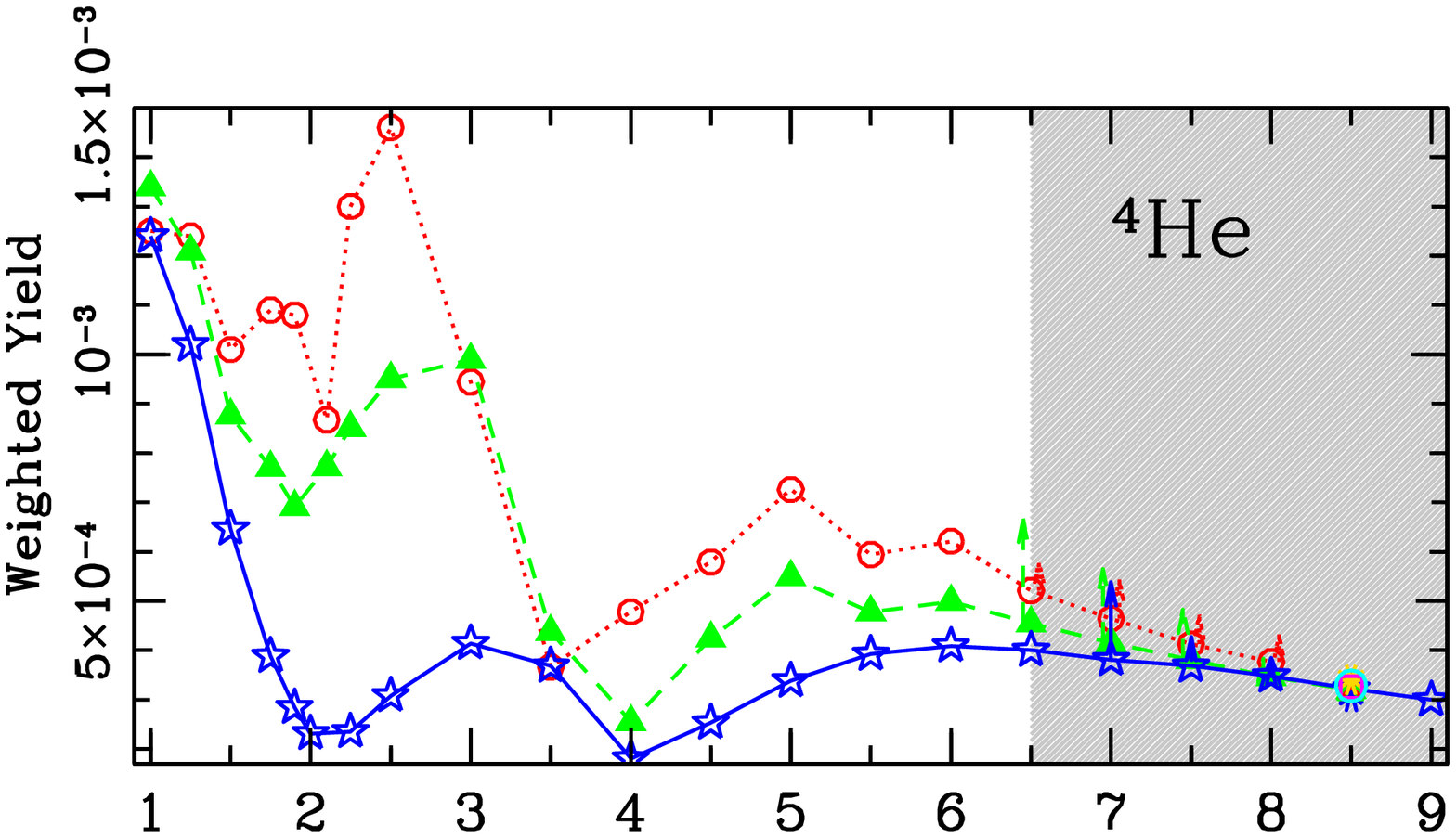}
 \includegraphics[width=8.5cm,angle=0]{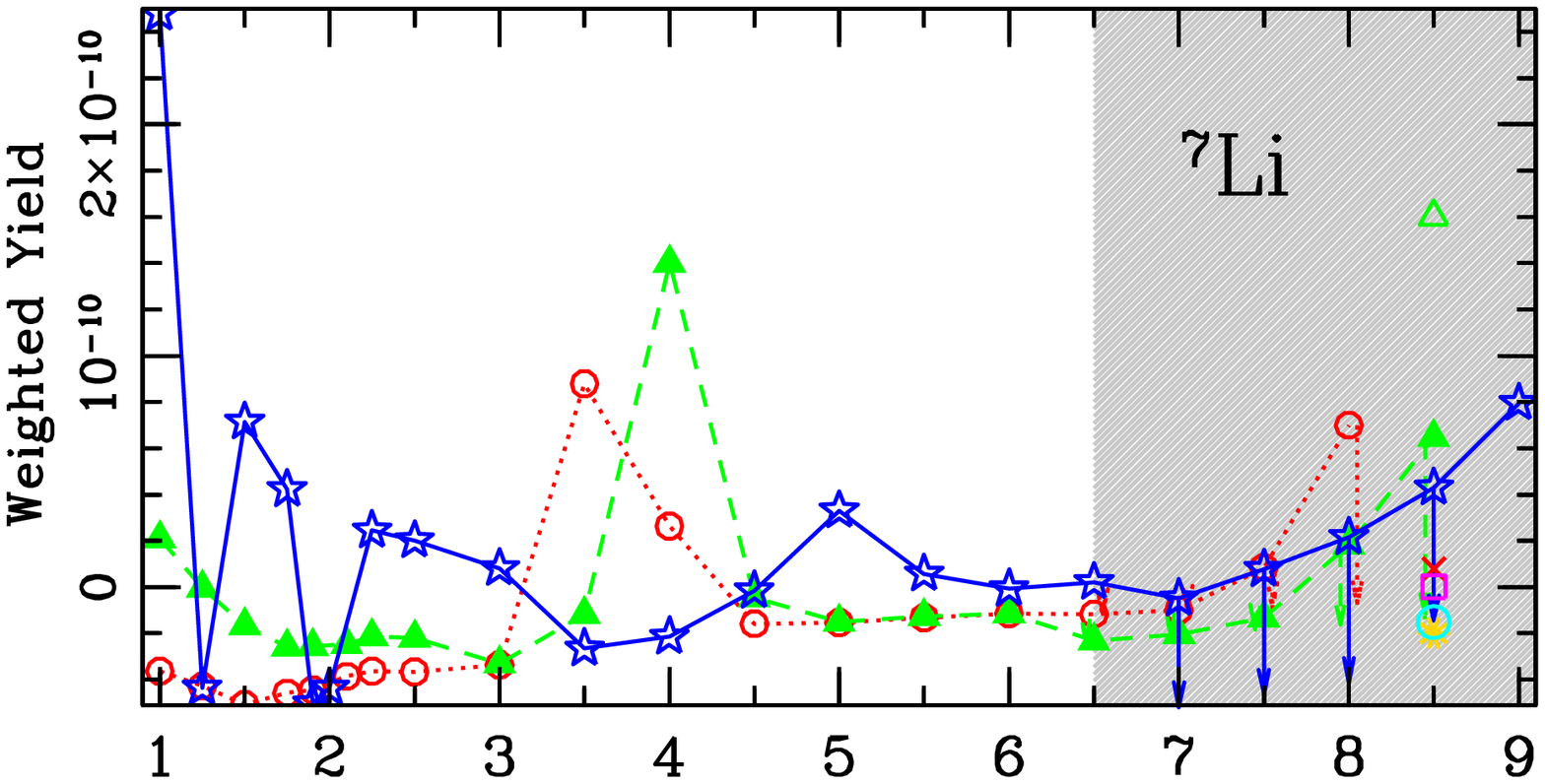}
 \includegraphics[width=8.5cm,angle=0]{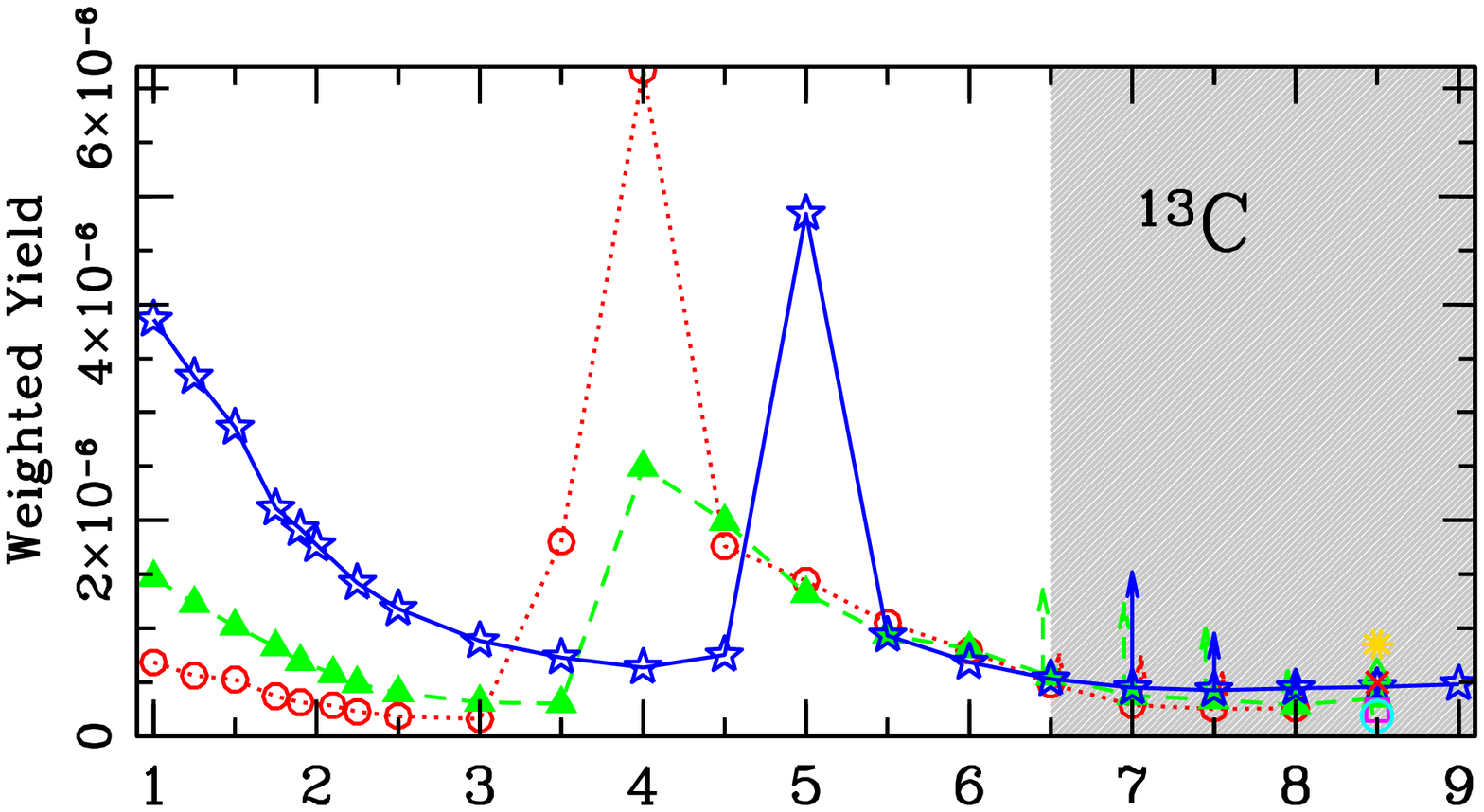}
 \includegraphics[width=8.5cm,angle=0]{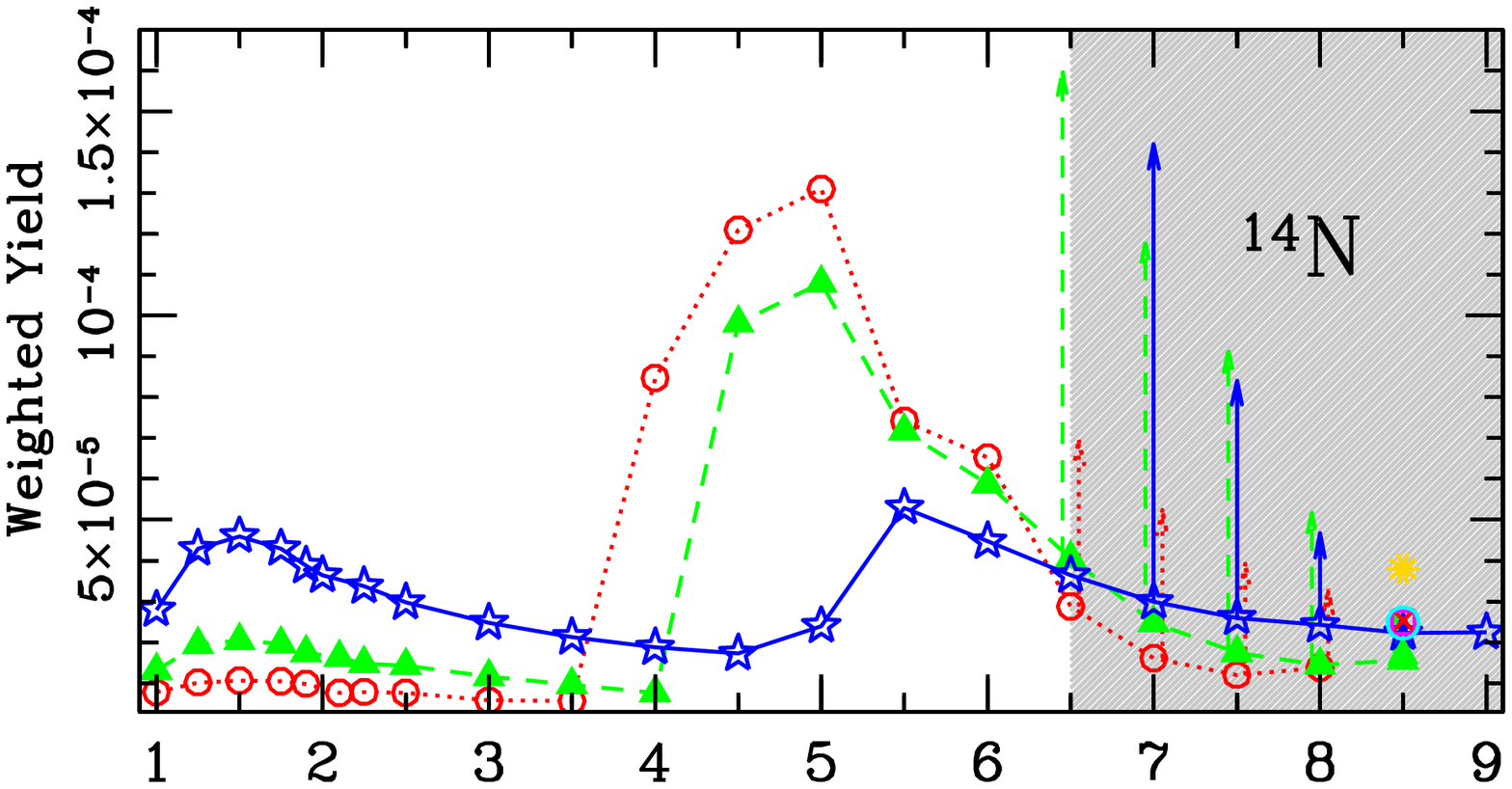}
 \includegraphics[width=8.5cm,angle=0]{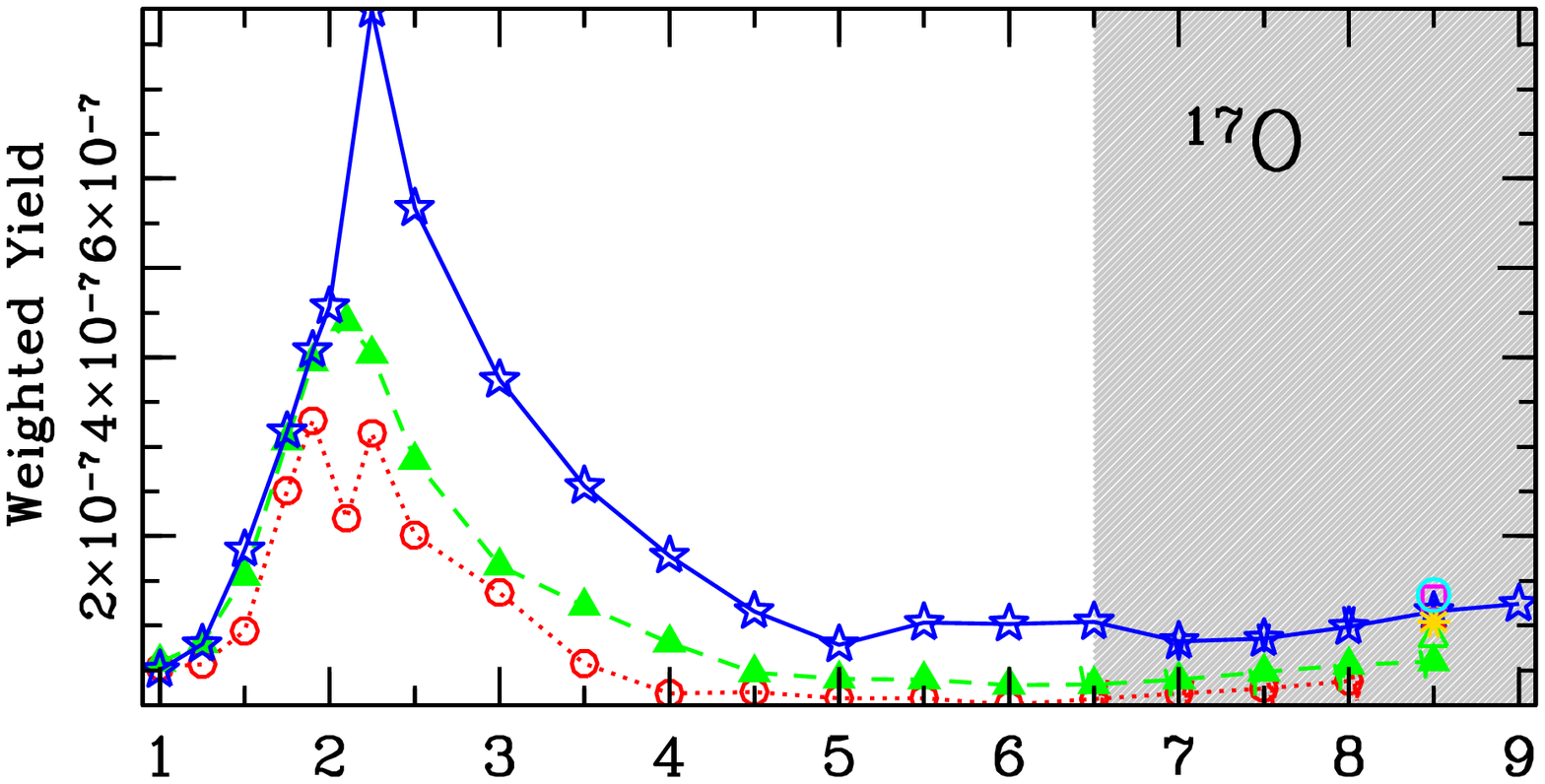}
 \includegraphics[width=8.5cm,angle=0]{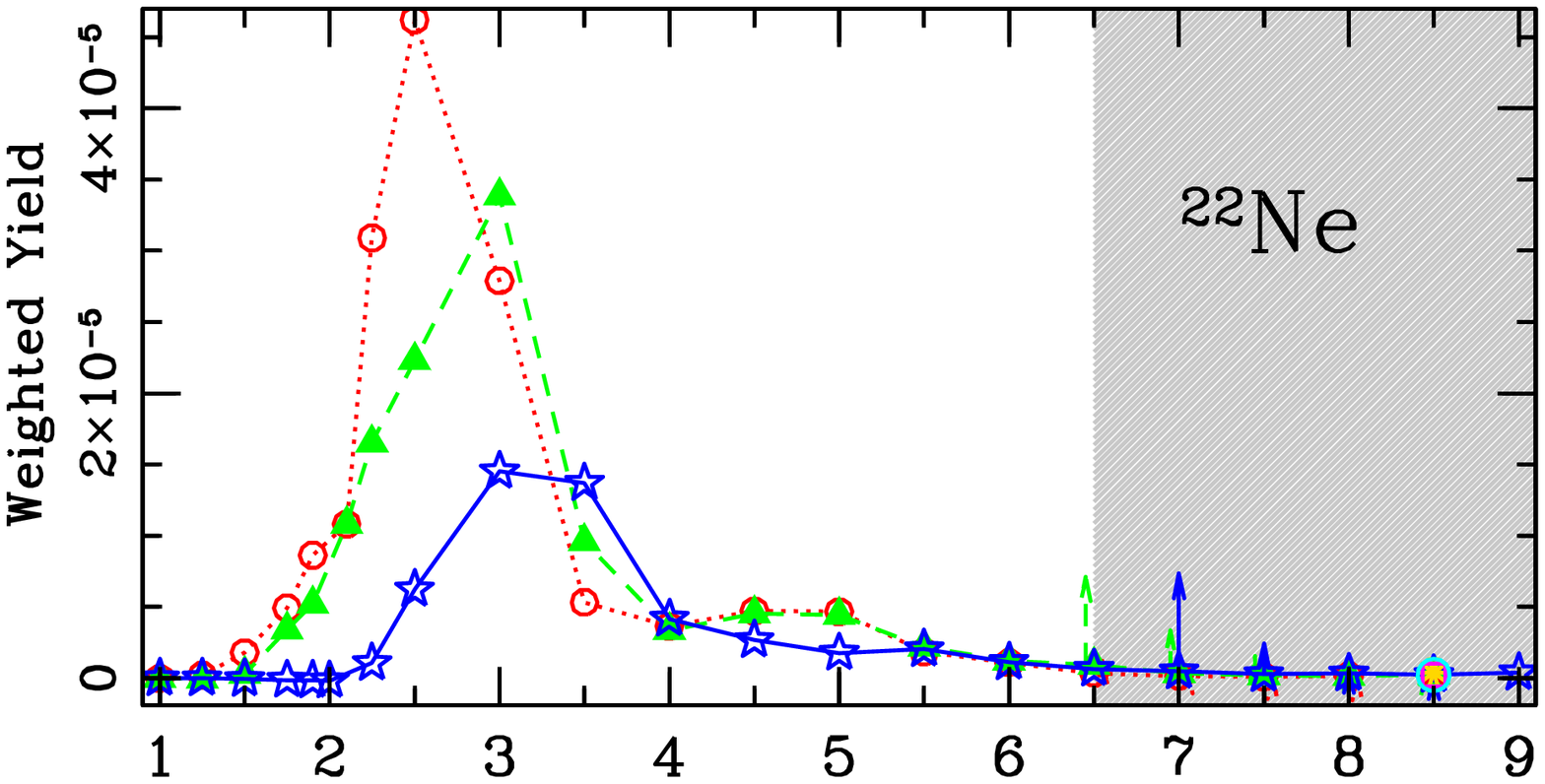}
 \includegraphics[width=8.5cm,angle=0]{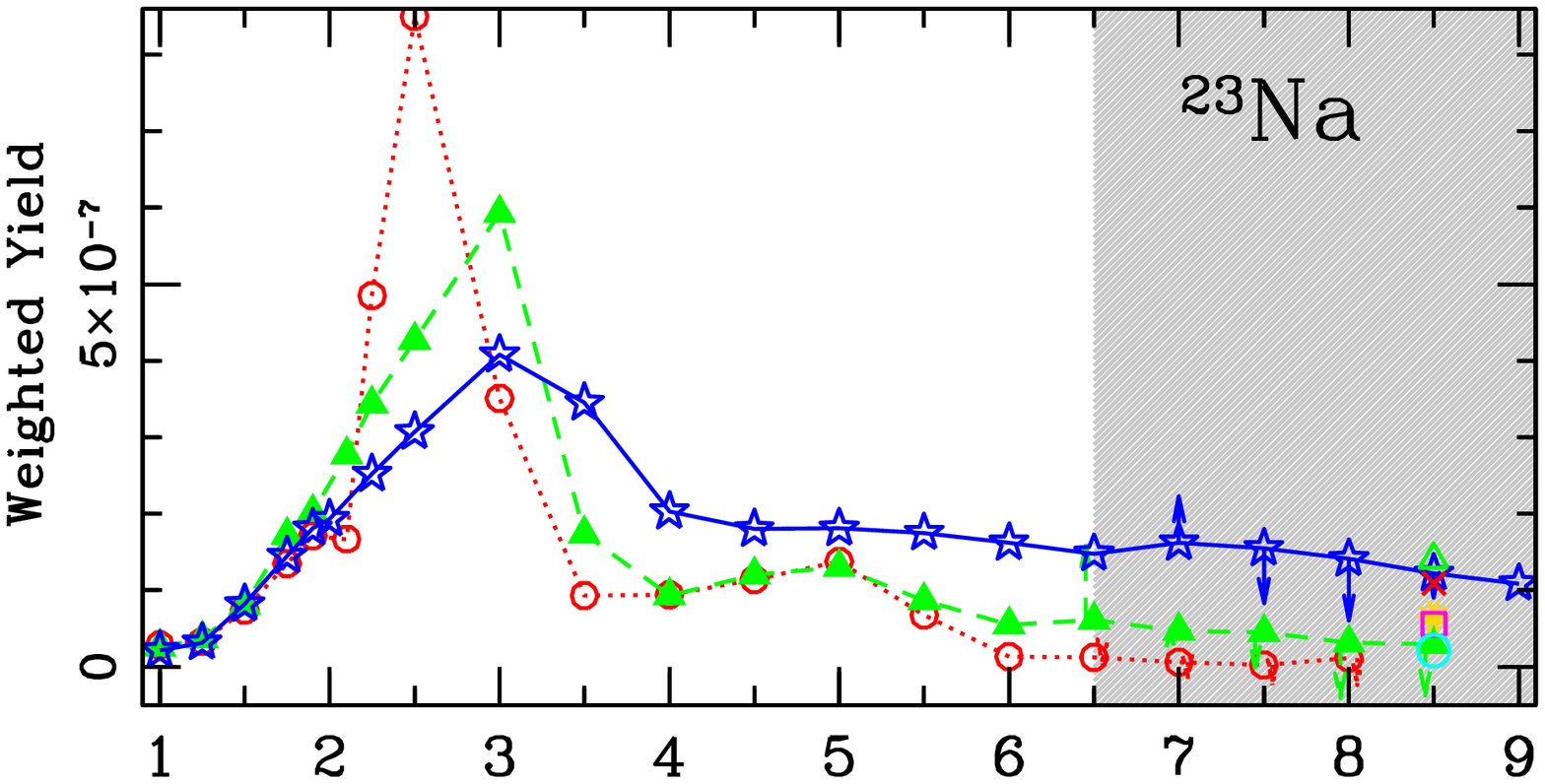}
 \includegraphics[width=8.5cm,angle=0]{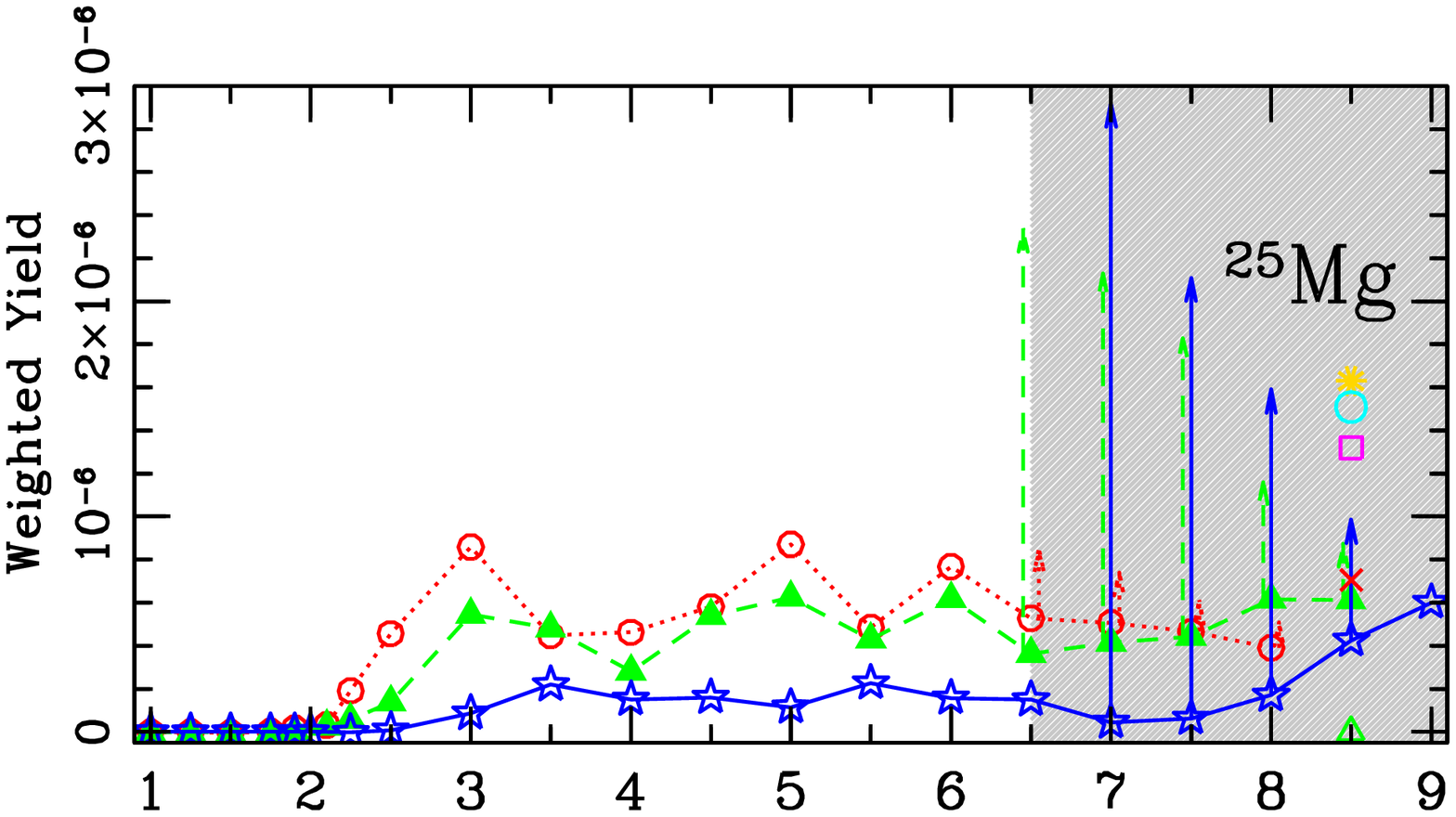}
 \includegraphics[width=8.5cm,angle=0]{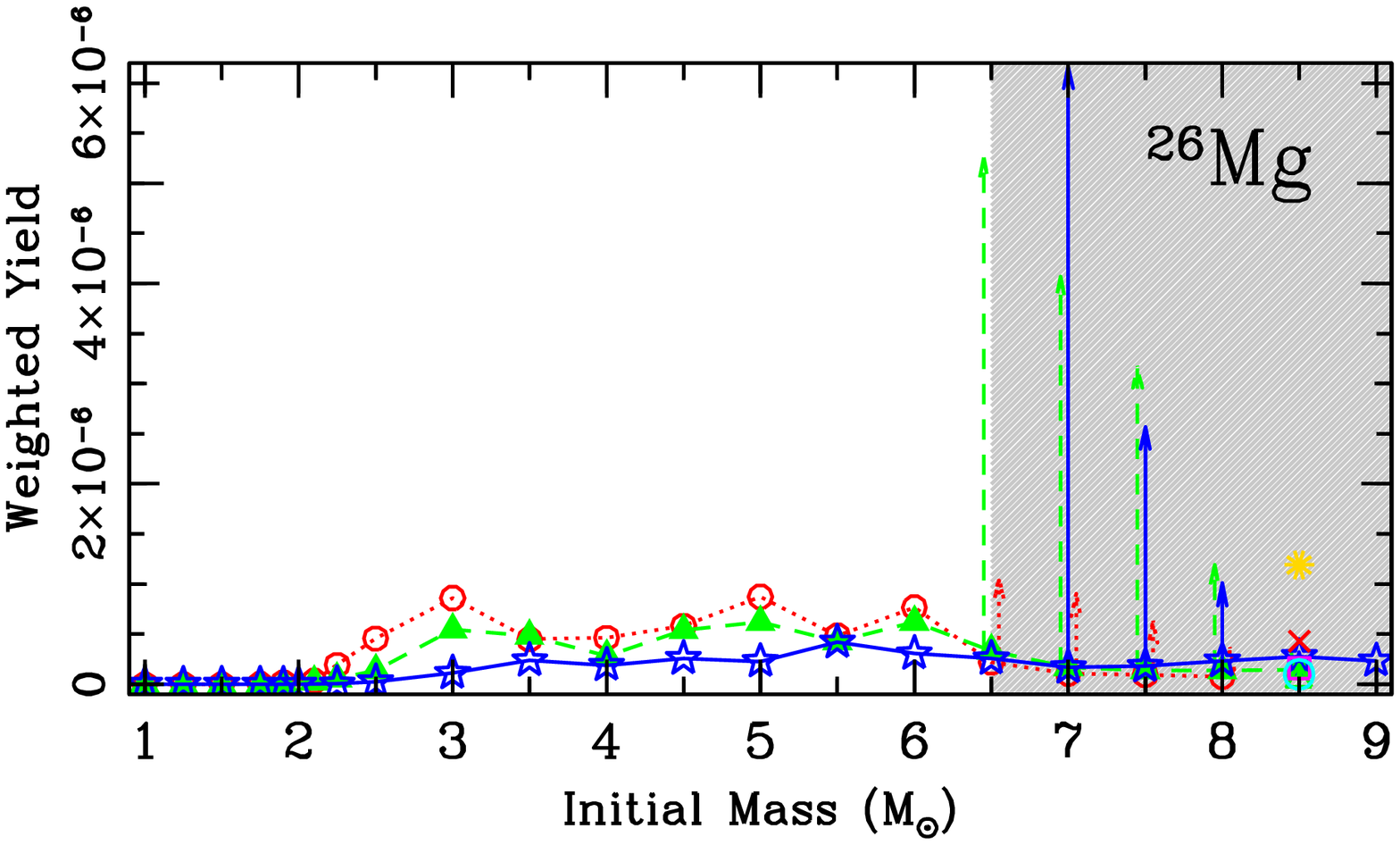}
 \includegraphics[width=8.5cm,angle=0]{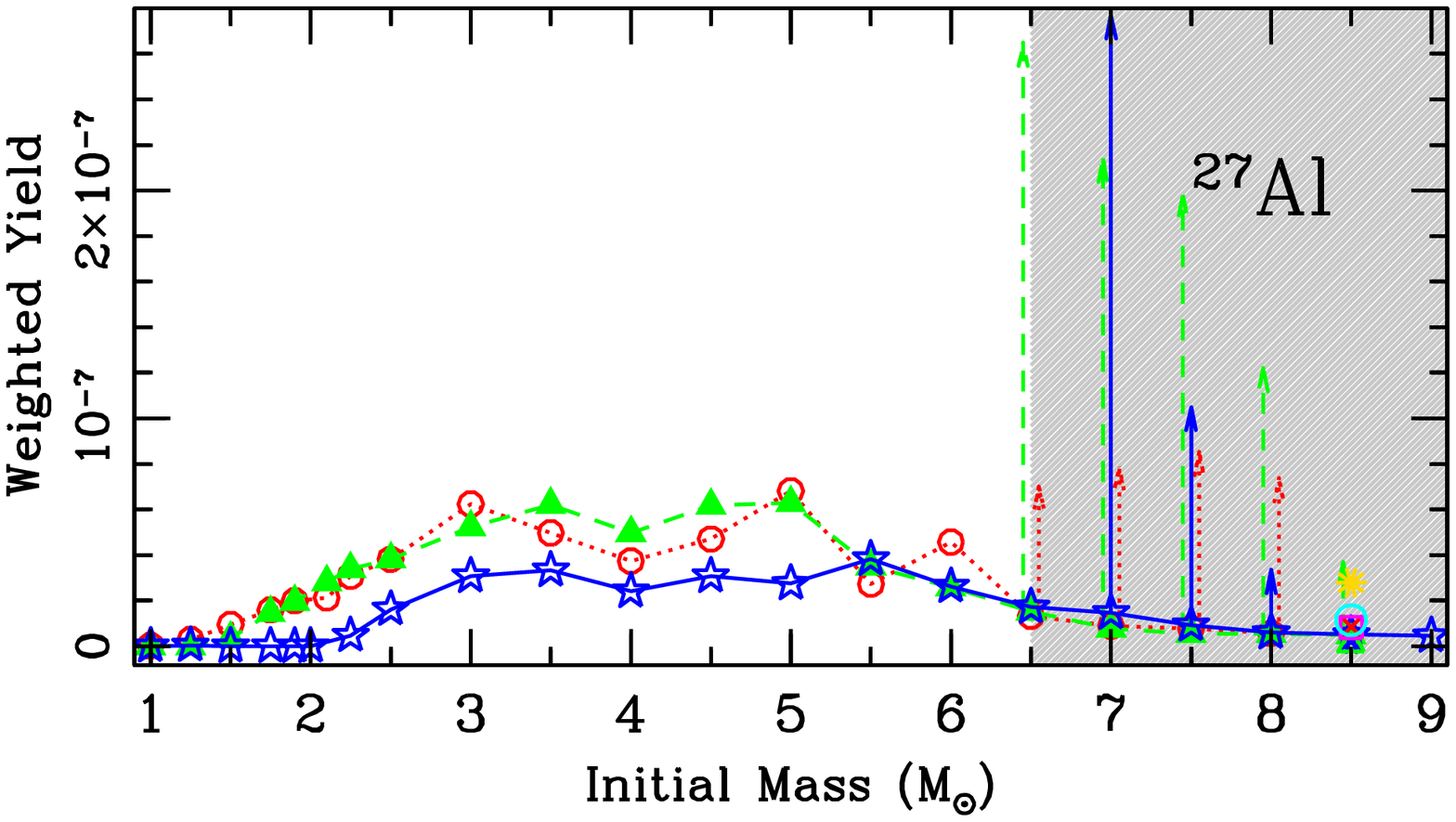}
 \caption{IMF (Kroupa) weighted yields (a non-dimensional quantity). The shaded regions represent the range of the models presented here. The lower mass AGB yields are from Karakas (2010). Calculations for Z=0.02 are shown with open stars and solid line, Z=0.008 with filled triangles and dashed line whilst Z=0.004 models are represented by open circles with dotted line. The test cases for 8.5 M$_\odot$ Z=0.02 are also shown, with symbols as per Figure~\ref{testcase}. The arrows represent the change due to the use of the delayed super-wind phase for the mass-loss rate.}
\label{fig-giants}
\end{center}
\end{figure*}

In Figure~\ref{fig-giants} we provide yields weighted by a \cite{kro93} initial mass function (IMF) of selected species for all standard and test case simulations.  
We weight the yields via the expression

\begin{equation}
y_{i} = \frac{dY}{dM} = \xi(\textrm{M}_{\rm{ini}}) M_{i}
\end{equation}

where $\xi$($M_{\rm{ini}}$)=0.15571 $M_{\rm{ini}}^{-2.7}$ for $M_{\rm{ini}}$ $>$ 1 M$_\odot$ \citep{hur02}, and $M_{i}$ is the net yield of species $i$.

The selected species are chosen from those most produced in this mass range as seen in Figure~\ref{productionfactor}.
The lower mass yield results are taken from \cite{kar10} which were produced using the \texttt{MONSTAR} program with input physics very similar to this work. These two sets of yields can be used in conjunction for galactic chemical evolution modelling.
Clearly seen in this figure is the relatively small contribution from super and massive AGB star yields of \chem{4}He, \chem{13}C, \chem{17}O, \chem{22}Ne and \chem{23}Na. This is irrespective of mass-loss rate. 
 The \chem{7}Li however could have a relatively large contribution but it strongly depends on the mass loss prescription. The contribution from \chem{14}N may also be important, with slow mass loss rate at the higher metallicities.
We expect super AGB stars may make quite a large contribution to the galactic inventory of the heavy magnesium isotopes and \chem{27}Al.
In most isotopes, apart from \chem{7}Li, \chem{17}O, and possibly \chem{25}Mg, the massive AGB stars contribution outweighs that of the super AGB stars.

\subsection{Convergence issues - extrapolated pulses}\label{ref-extrap}
As seen in all previous studies of intermediate-mass AGB stellar models the evolution is terminated due to convergence issues prior to removal of the entire envelope, with the most massive models retaining up to $\approx$ 2.5 M$_\odot$ of envelope. In \cite{lau12} these convergence issues were attributed to the iron opacity peak.
Although we expect the evolution to be truncated due to this instability with possible expulsion of the entire remaining envelope in one large mass ejection event, the outcome is uncertain, with a possibility that the stars may recover stability after this hydrodynamical event and undergo additional thermal pulses. Therefore we have decided to also model synthetically the possible remaining thermal pulses in the method described in \cite{kar03}.

Our standard set of yield calculations assume the remaining envelope is expelled with the composition it had when convergence issues ceased calculations. Synthetic calculations at the end of the evolution can provide an approximation of the uncertainties in these yields.

For the extrapolated pulses we take input values of $\lambda$ and the interpulse period $\tau_{\rm{IP}}$ from the final calculated pulse, whilst for the values of radius, luminosity, and $T_{\rm{eff}}$ we take the average of the values during the last interpulse period. The intershell composition is taken from the last computed thermal pulse, noting that the temperature is very similar between sequential TPs nearing the end of the evolution. The HBS is also very thin in mass compared to the intershell, so the assumption that the 3DU is composed only of intershell material is a good approximation. The increase in core mass during these final synthetic calculations is at most $\sim$ 0.01 M$_\odot$ (8.5 M$_\odot$ Z=0.02 R75) and is taken into account when calculating the total mass expelled.

In Tables~\ref{tpsagb} and \ref{testcasetable} the values in the square brackets are the number of extrapolated pulses. For the test cases we have used the appropriate mass loss rate also for the synthetic calculations.
We find that by the time the envelope instability develops HBB has already ceased in all models.
As a consequence, the subsequent evolution of the envelope composition is only due to the 3DU pollution (mainly of \chem{4}He, \chem{12}C, \chem{16}O and to a lesser extent \chem{25}Mg, \chem{26}Mg, \chem{19}F, \chem{21}Ne and \chem{22}Ne) which leaves \chem{13}C and \chem{14}N almost unchanged.

\begin{figure}
\resizebox{\hsize}{!}{\includegraphics{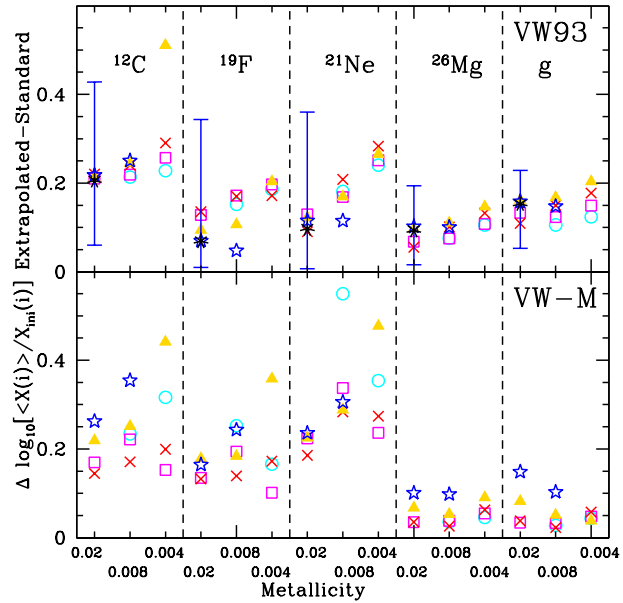}}
\caption{Difference between standard yield production factors and those which include the effects of extrapolated pulses for VW93 (top panel) and VW-M (bottom panel) models. Colors and symbols as per Figure~\ref{productionfactor}. The error bars on the 8.5 M$_\odot$ Z=0.02 model encompass the range of values found for the test cases.}
\label{fig-extrap}
\end{figure}

Because we do not take into account the possible increase in mass-loss rate and/or decrease in 3DU efficiency with decreasing envelope mass, we most likely over-estimate the contribution from the extrapolated TPs to the stellar yields.
Generally, the stellar models most affected by the addition of extrapolated pulses are those of lower metallicity, due to the comparatively larger enrichment from their lower initial abundances. The difference in standard compared to extrapolated production factors are seen in Figure~\ref{fig-extrap}. The species chosen for this figure are those which show the largest variations due to the extrapolated thermal pulses.

In the yield calculations with the contribution from the synthetic pulses, there is at most a $\sim$ 0.2 dex increase in \chem{26}Mg and g. However, the \chem{12}C, and also the less abundant isotopes \chem{19}F and \chem{21}Ne show quite large enrichments of up to $\sim$ 0.5 dex. Even including the extrapolated pulses, the \chem{19}F and \chem{21}Ne yields are still strongly negative. The \chem{12}C yields can change from negative to positive when the extrapolations are included, with the possible 0.5 dex increase in these yields representing one of the largest uncertainties for this isotope. The error bars on the 8.5 M$_\odot$ Z=0.02 represent the bounds of the mass-loss rate and varying mixing length tests, with the largest/smallest contributions from the R75 and B95 models respectively.
The set of yield calculations including extrapolated pulses can be found in the online tables.

 \subsection{Comparison with other results}

\begin{figure*}
\begin{center}
\resizebox{\hsize}{!}{\includegraphics{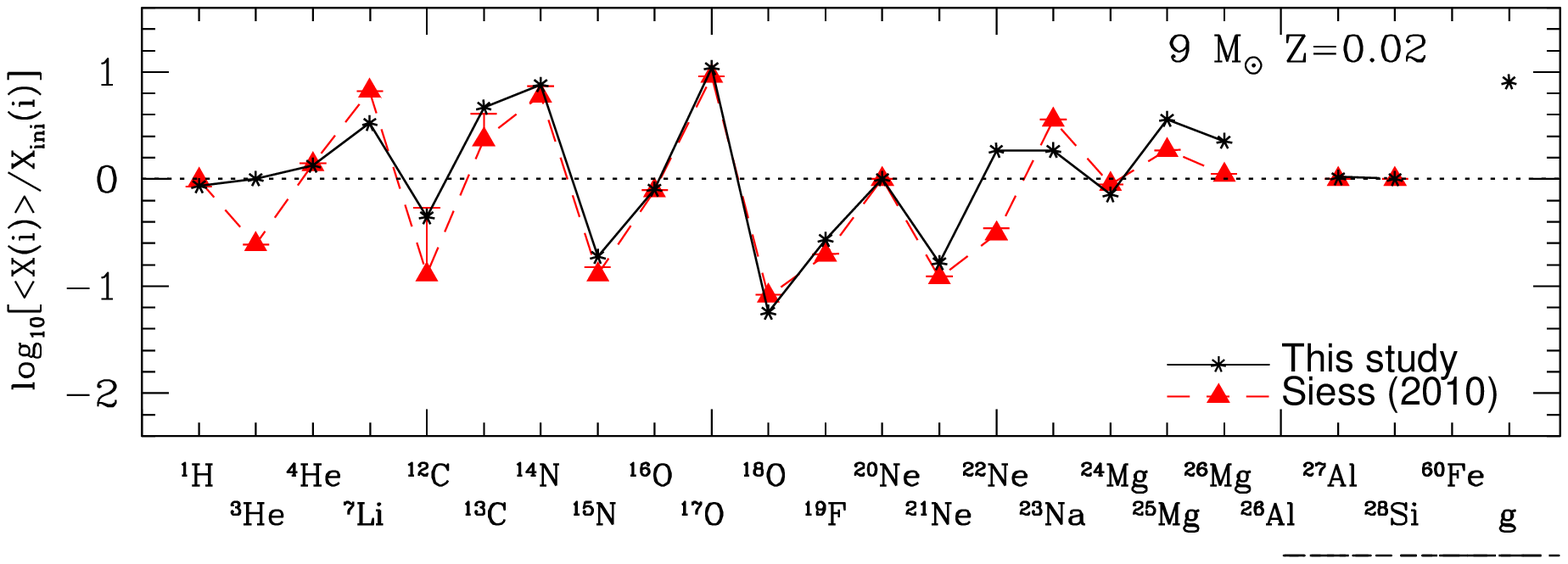}}
\resizebox{\hsize}{!}{\includegraphics{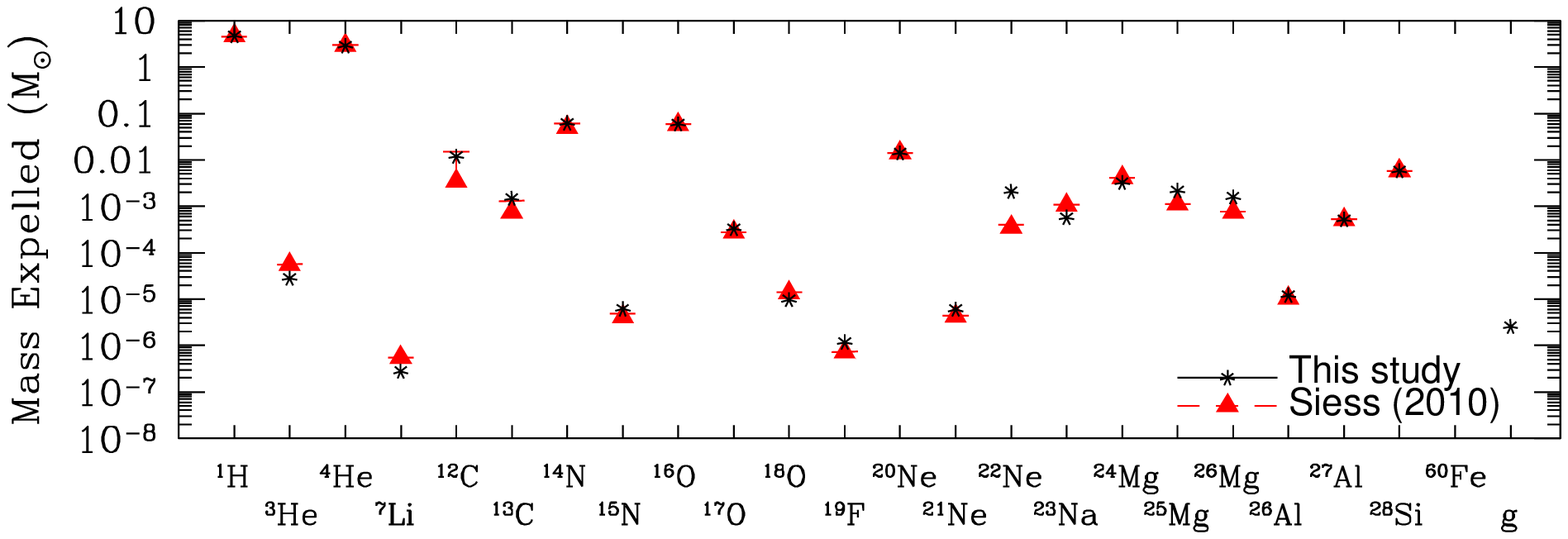}}
\caption{Production factor (top panel) and the total mass expelled (M$_\odot$) (bottom panel) for the 9 M$_\odot$ Z=0.02 model from this study compared to that from Siess (2010). The error bar on S10 results include his synthetic 3DU calculations with $\lambda$=0.8. Both sets of models have included extrapolated pulses.}
\label{compare}\end{center}
\end{figure*}

Only \cite{sie10}, hereafter S10, has produced stellar nucleosynthesis and yields for the entire metallicity and mass range of this study.\footnote{The super AGB stars models of \cite{ven10a,ven11b,ven13} and \cite{her12} are for the globular cluster metallicities and comparison to these studies is deferred to paper IV.} 
We compare with S10 the only model with mass and metallicity common to both studies: the 9 M$_\odot$ Z=0.02 model (with extrapolated pulses).
Figure~\ref{compare} compares the models from both studies for a variety of isotopes. The top panel includes the production factor values, whilst the bottom panel shows the total mass expelled for a selection of species. In Table~\ref{compare2} we collect a selection of model characteristics.

There are two major differences between the results from S10 and this work. The first is the occurrence of 3DU in this work, whilst the second is the quite large difference in the core mass, with the models presented here having more massive cores. The latter is due to different treatment of convective boundaries during core helium burning.

As the temperature at the base of the convective envelope is strongly dependent on, and is an increasing function of, core mass, the S10 model with its less massive core for the same initial mass has resulted in the lower HBB temperature.
There is also a variation in the number of thermal pulses, 340 in S10 compared to 297 in this work. The almost halving of the duration of the TP-(S)AGB phase in our model is a result of higher luminosity and the more rapid mass-loss rate associated with its larger core mass.  
S10 has also produced synthetic stellar yields which artificially include the effects of 3DU, or mimic the effects of increasing $\alpha_{\rm{mlt}}$ via an increase in $T_{\rm{BCE}}$. We compare to his 3DU model with $\lambda$ = 0.8 and include this in Figure~\ref{compare} as error-bars. 
Third dredge-up events have substantial feedback on the evolution, via increased cooling of the intershell regions. This will lead to a substantial increase in the interpulse period  \cite[e.g.][]{sac77,fro96,sta04} and subsequent divergence in number of TPs for the same thermally pulsing phase duration.
Even with these large physical differences between the code inputs we note surprisingly close agreement between the model results as seen in Figure~\ref{compare}. 
This is because, at least for the metal rich models provided here, the effects of 3DU are much out-weighted by hot and very efficient HBB.   

From Figure~\ref{compare} the main differences between these results
primarily concern \chem{22}Ne, \chem{23}Na, \chem{25}Mg and \chem{26}Mg. The large \chem{22}Ne abundance in our models is a direct consequence of both the corrosive 2DU and 3DU episodes and the lower \chem{23}Na yields can be explained by our use of a slower \chem{22}Ne(p,$\gamma$) rate (\cite{hal04} in this work versus NACRE in S10). 
The heavy magnesium isotope differences are due to both the higher $T_{\rm{BCE}}$ for \chem{25}Mg and the occurrence of 3DU for \chem{25}Mg and \chem{26}Mg. Another obvious signature of 3DU which is absent from S10 is the production of s-process neutron capture elements in our study.
The difference in lithium yield is due to the lower $T_{\rm{BCE}}$ in the S10 model, whilst the \chem{3}He variations is mostly due to the adopted initial compositions.
\chem{12}C and \chem{13}C yields show very good agreement when compared to the $\lambda$ = 0.8 yields from S10.

\begin{table}
\begin{center}\setlength{\tabcolsep}{1.5pt} 
\caption{Comparison of selected model characteristics for 9 M$_\odot$ Z=0.02 from this study and Siess (2010). Variables as described in Table~\ref{tpsagb}.}
\label{compare2}
\begin{tabular}{lcccccrc} \hline \hline
&$T_{\rm{BCE}}^{\rm{Max}}$&$M_{\rm{Dredge}}^{\rm{Tot}}$ & $M_{\rm{2DU}}$ & $M_{\rm{C}}^{\rm{F}}$ &$M_{\rm{env}}^{\rm{F}}$ &  $N_{\rm{TP}}$  & $\tau_{\rm{(S)AGB}}$ \\
&(MK)&(M$_\odot$)&(M$_\odot$)&(M$_\odot$)&(M$_\odot$)&&(yrs)\\\hline
 \multicolumn{8}{c}{9 M$_\odot$ Z=0.02} \\ \hline
This Study  &113&5.08(-2)&1.21 & 1.23 &1.82 &221[76]  & 9.24(4) \\
Siess (2010)&99 &0       &1.06 & 1.15 &2.85 &119[221] & 1.71(5)\\
\hline
 \end{tabular} \end{center} \end{table}

\section{Summary and Conclusions}

The stellar yields produced in this study confirm that super and massive AGB stars in the metallicity range $0.004 \la Z \la 0.02$ are large producers of \chem{4}He, \chem{13}C, \chem{14}N, \chem{17}O, \chem{23}Na, \chem{25}Mg and \chem{26}Al. These models with their efficient 3DU also produce considerable amounts of \chem{22}Ne, \chem{26}Mg and \chem{60}Fe.
However from a galactic perspective that takes into account the mass distribution of super AGB stars, we find that these stars mainly contribute to the ISM enrichment in \chem{25}Mg, \chem{26}Mg (and possibly \chem{14}N, \chem{7}Li and \chem{27}Al). 

In our selected metallicity range, the surface abundance contribution from 3DU is clearly outweighed by the effects of hot bottom burning. 

We have explored a range of uncertainties in our study of super and massive AGB star nucleosynthesis. The effect of mass-loss changes were found to mostly impact on the amount of third dredge-up enrichment, with this more important in massive AGB stars compared to in super AGB stars. The extrapolations used to deal with evolution after convergence issues contribute to the uncertainties in only a few select species, mainly \chem{12}C, \chem{19}F, \chem{21}Ne, \chem{26}Mg and g. The yield of \chem{7}Li was found to be notoriously difficult to constrain because of its fragility, with its yield highly dependent on both the mass-loss rate and choice of mixing-length parameter.

In the search for a nucleosynthetic signature of a super AGB star that would distinguish it from a massive AGB, we could not identify a clear candidate, at least for elements lighter than iron.
Results for heavier than iron (s-process) element nucleosynthetic yields for super AGB stars using the \texttt{MONTAGE} post processing code \citep{chu09} will be presented in Lau et al. (2013, in preparation). 

\section{Acknowledgments}
This research was supported under Australian Research Council's Discovery Projects funding scheme (project number DP0877317). CLD would like to thank J Ortiz Domenech and G Kennedy for their wonderful hospitality during her extended stays in Barcelona.

\bibliographystyle{mn2e}

\bibliography{mnras}

\appendix

\section{Online material} We include supplementary electronic online tables as follows\\
{\bf{Table~1.}} Stellar Yields for Z=0.02, 0.008 and 0.004 with VW93 mass-loss rate.\\
{\bf{Table~2.}} Stellar Yields for Z=0.02, 0.008 and 0.004 with VW-M mass-loss rate.\\
{\bf{Table~3.}} Stellar Yields for 8.5 M$_\odot$ Z=0.02 test cases.\\
{\bf{Table~4.}} Initial Composition in mass fraction for  Z=0.02, 0.008 and 0.004.\\

\section{Supplementary Tables}
\begin{table*}
\begin{center} 
\caption{Initial, First and Second Dredge-up surface abundances of selected isotopes for our standard VW93 models for metallicities Z=0.02, 0.008 and 0.004. Where $n(m)= n\times 10^{m}$.}
\label{12du} \setlength{\tabcolsep}{3pt} 
\begin{tabular}{lccccccccccccccc} \hline \hline
$M_{\rm{ini}}$& Phase&\chem{1}H&\chem{4}He&\chem{12}C&\chem{13}C&\chem{14}N&\chem{15}N&\chem{16}O&\chem{17}O&\chem{18}O&\chem{21}Ne&\chem{22}Ne&\chem{23}Na&\chem{25}Mg&\chem{26}Mg\\ \hline
\hline \multicolumn{16}{c}{Z=0.02} \\ \hline
- &Initial&0.705&0.275&3.42(-3)&4.11(-5)&1.05(-3)&4.15(-6)&9.52(-3)&3.88(-6)&2.16(-5)&4.62(-6)&1.45(-4)&3.98(-5)&7.69(-5)&8.81(-5)\\ \hline
7&1DU &0.696&0.284&2.14(-3)&1.05(-4)&2.99(-3)&1.87(-6)&8.93(-3)&1.39(-5)&1.53(-5)&5.03(-6)&1.28(-4)&5.83(-5)&7.52(-5)&8.99(-5)\\
7 &2DU    &0.628&0.352&1.89(-3)&9.81(-5)&4.05(-3)&1.63(-6)&8.06(-3)&1.33(-5)&1.37(-5)&8.38(-6)&1.14(-4)&7.34(-5)&6.80(-5)&9.70(-5)\\
7.5 & 1DU &0.695&0.285&2.12(-3)&1.05(-4)&3.07(-3)&1.84(-6)&8.86(-3)&1.35(-5)&1.51(-5)&5.12(-6)&
1.27(-4)&5.93(-5)&7.48(-5)&9.03(-5)\\
7.5&2DU&0.622&0.358&1.86(-3)&9.74(-5)&4.20(-3)&1.59(-6)&7.93(-3)&1.28(-5)&1.82(-5)&8.33(-6)&1.13(-4)&7.63(-5)&6.71(-5)&9.79(-5)\\
8&1DU&0.694&0.286&2.11(-3)&1.05(-4)&3.12(-3)&1.83(-6)&8.82(-3)&1.32(-5)&1.51(-5)&5.18(-6)&1.26(-4)&5.99(-5)&7.45(-5)&9.06(-5)\\
8&2DU& 0.617&0.363&1.90(-3)&9.71(-5)&4.27(-3)&1.56(-6)&7.84(-3)&1.24(-5)&3.25(-5)&8.06(-6)&1.57(-4)&7.85(-5)&6.63(-5)&9.85(-5)\\
8.5&1DU&0.692&0.287&2.10(-3)&1.05(-4)&3.19(-3)&1.81(-6)&8.76(-3)&1.28(-5)&1.49(-5)&5.27(-6)&1.25(-4)&6.09(-5)&7.41(-5)&9.10(-5)\\
8.5&2DU&0.614&0.366&1.92(-3)&1.01(-4)&4.35(-3)&9.57(-7)&7.76(-3)&1.20(-5)&3.51(-5)&7.69(-6)&1.71(-4)&8.06(-5)&6.57(-5)&9.90(-5)\\
9&1DU&0.692&0.288&2.09(-3)&1.05(-4)&3.23(-3)&1.80(-6)&8.73(-3)&1.25(-5)&1.49(-5)&5.33(-6)&1.25(-4)&6.13(-5)&7.39(-5)&9.13(-5)\\
9&2DU&0.610&0.369&2.67(-3)&1.42(-4)&4.41(-3)&1.05(-6)&7.88(-3)&1.18(-5)&3.14(-5)&7.13(-6)&2.47(-4)&8.25(-5)&6.56(-5)&9.98(-5)\\
\hline \multicolumn{16}{c}{Z=0.008} \\ \hline
- &Initial&0.733&0.259&1.42(-3)&1.64(-5)&4.36(-4)&1.66(-6)&3.96(-3)&1.55(-6)&8.64(-6)&1.85(-6)&5.81(-5)&1.59(-5)&3.08(-5)&3.52(-5)\\ \hline
6.5&1DU&0.732&0.259&9.08(-4)&4.51(-5)&1.05(-3)&7.55(-7)&3.91(-3)&6.70(-6)&6.31(-6)&1.87(-6)&5.39(-5)&2.03(-5)&3.07(-5)&3.53(-5)\\
6.5&2DU&0.658&0.333&7.85(-4)&3.97(-5)&1.66(-3)&6.63(-7)&3.39(-3)&6.42(-6)&5.42(-6)&3.80(-6)&4.64(-5)&3.10(-5)&2.71(-5)&3.89(-5)\\
7&1DU&0.732&0.260&9.02(-4)&4.50(-5)&1.07(-3)&7.47(-7)&3.89(-3)&6.75(-6)&6.25(-6)&1.89(-6)&5.35(-5)&2.07(-5)&3.06(-5)&3.54(-5)\\
7&2DU&0.651&0.341&7.77(-4)&3.94(-5)&1.71(-3)&6.50(-7)&3.34(-3)&6.25(-6)&6.17(-6)&3.35(-6)&4.58(-5)&3.29(-5)&2.67(-5)&3.92(-5)\\
7.5&1DU&0.731&0.260&8.92(-4)&4.48(-5)&1.12(-3)&7.36(-7)&3.85(-3)&6.74(-6)&6.17(-6)&1.92(-6)&5.29(-5)&2.14(-5)&3.04(-5)&3.56(-5)\\
7.5&2DU&0.646&0.346&8.22(-4)&3.93(-5)&1.74(-3)&6.38(-7)&3.31(-3)&6.04(-6)&1.30(-5)&2.98(-6)&5.90(-5)&3.44(-5)&2.64(-5)&3.95(-5)\\
8&1DU&0.731&0.261&8.85(-4)&4.46(-5)&1.16(-3)&7.30(-7)&3.82(-3)&6.62(-6)&6.11(-6)&1.96(-6)&5.24(-5)&2.19(-5)&3.03(-5)&3.58(-5)\\
8&2DU&0.644&0.348&8.83(-4)&4.39(-5)&1.75(-3)&3.13(-7)&3.29(-3)&5.83(-6)&1.45(-5)&2.67(-6)&7.22(-5)&3.54(-5)&2.62(-5)&3.96(-5)\\
8.5&1DU&0730&0.262&8.74(-4)&4.45(-5)&1.21(-3)&7.18(-7)&3.78(-3)&6.48(-6)&6.03(-6)&2.02(-6)&5.18(-5)&2.25(-5)&3.00(-5)&3.60(-5)\\
8.5&2DU&0.640&0.351&1.73(-3)&1.04(-4)&1.83(-3)&3.46(-7)&3.43(-3)&5.89(-6)&1.11(-5)&2.32(-6)&1.12(-4)&3.66(-5)&2.61(-5)&3.99(-5)\\
\hline \multicolumn{16}{c}{Z=0.004} \\ \hline
- &Initial& 0.743&0.253&7.20(-4)&8.21(-6)&2.21(-4)&8.29(-7)&2.01(-3)&7.75(-7)&4.32(-6)&9.23(-7)&
2.91(-5)&7.96(-6)&1.54(-5)&1.76(-5)\\\hline
6.5&1DU&0.742&0.253&4.93(-4)&2.48(-5)&4.69(-4)&4.00(-7)&2.01(-3)&1.50(-6)&3.40(-6)&9.24(-7)&2.84(-5)&8.60(-6)&1.54(-5)&1.76(-5)\\
6.5&2DU&0.653&0.342&3.81(-4)&1.95(-5)&8.79(-4)&3.13(-7)&1.70(-3)&4.44(-6)&3.34(-6)&1.25(-6)&2.25(-5)&2.10(-5)&1.32(-5)&1.95(-5)\\
7&1DU&0.742&0.253&4.86(-4)&2.46(-5)&4.78(-4)&3.92(-7)&2.01(-3)&1.72(-6)&3.34(-6)&9.24(-7)&2.83(-5)&8.78(-6)&1.54(-5)&1.76(-5)\\
7&2DU&0.649&0.347&4.28(-4)&1.94(-5)&8.93(-4)&3.08(-7)&1.68(-3)&4.37(-6)&6.99(-6)&1.16(-6)&3.05(-5)&2.19(-5)&1.31(-5)&1.96(-5)\\
7.5&1DU&0.742&0.253&4.76(-4)&2.42(-5)&4.90(-4)&3.83(-7)&2.01(-3)&2.09(-6)&3.27(-6)&9.24(-7)&2.80(-5)&9.06(-6)&1.54(-5)&1.76(-5)\\
7.5&2DU&0.646&0.349&5.44(-4)&2.37(-5)&9.02(-4)&1.67(-7)&1.67(-3)&4.23(-6)&7.54(-6)&1.11(-6)&3.93(-5)&2.25(-5)&1.30(-5)&1.96(-5)\\
8&1DU&0.742&0.253&4.69(-4)&2.40(-5)&4.99(-4)&3.76(-7)&2.00(-3)&2.49(-6)&3.22(-6)&9.24(-7)&2.77(-5)&9.33(-6)&1.54(-5)&1.76(-5)\\
8&2DU&0.642&0.352&1.50(-3)&8.43(-5)&9.68(-4)&1.55(-7)&1.97(-3)&4.20(-6)&5.23(-6)&1.04(-6)&5.58(-5)&2.32(-5)&1.31(-5)&1.99(-5)\\
\hline 
\end{tabular}
 \end{center}
 \end{table*}

\begin{table*}
\begin{center}\setlength{\tabcolsep}{3pt} 
\caption{Sample first few lines of yield table (in solar units).}\label{tableappend2}
\begin{tabular}{ccccccc}
\hline
\multicolumn{7}{c}{$M_{\rm{ini}}$ = 9 M$_{\odot}$ Z = 0.02} \\ 
\hline \hline 
&&&&\multicolumn{3}{c}{Extrapolated TPs}\\
Species $i$ & Net yield (M$_{\odot}$) & $M^{\mathrm{wind}}(i)$ (M$_{\odot}$) &$\log_{10}[\langle X(i)
  \rangle/X_{\mathrm{ini}}(i)]$ & Net yield (M$_{\odot}$) &$M^{\mathrm{wind}}(i)$ (M$_{\odot}$) &$\log_{10}[\langle X(i)\rangle/X_{\mathrm{ini}}(i)]$ \\
\hline
H          & -4.832E-01 &  3.770E+00 & -5.240E-02 & -4.927E-01 &  3.760E+00 & -5.350E-02\\
\chem{3}He &  1.589E-05 &  1.589E-05 &  0.000E+00 &  1.587E-05 &  1.587E-05 &  0.000E+00\\
\chem{4}He &  4.687E-01 &  2.128E+00 &  1.080E-01 &  4.736E-01 &  2.133E+00 &  1.090E-01\\
\chem{7}Li & -6.037E-09 &  5.857E-08 & -4.263E-02 & -6.088E-09 &  5.852E-08 & -4.301E-02\\
\chem{12}C & -1.386E-02 &  6.763E-03 & -4.843E-01 & -9.659E-03 &  1.096E-02 & -2.744E-01\\
\hline
\end{tabular}
\medskip\\
\end{center}
\end{table*}

\begin{table*}
\begin{center}
\setlength{\tabcolsep}{3pt} 
\caption{Selected model characteristics for the thermal pulsing (S)AGB phase including extrapolated thermal pulses. Variables as described in text.}\label{extrapstuff}
\begin{tabular}{lcccrccccrccc}
\hline \hline 
 & & & & && & & \multicolumn{5}{c}{Extrapolated TPs}\\
$M_{\rm{ini}}$ & C/O$_{\rm{2DU}}$ &$L^{\rm{Max}}$&  $M_{\rm{Dredge}}^{\rm{Tot}}$ & $N_{TP}$ & $\tau_{\rm{(S)AGB}}$ & C/O$_{\rm{F}}$ & $\tau_{\rm{C}}$/$\tau_{\rm{M}}$ & $M_{\rm{Dredge}}^{\rm{Tot}}$ &  $N_{TP}$& $\tau_{\rm{(S)AGB}}$  & C/O$_{\rm{F}}$ & $\tau_{\rm{C}}$/$\tau_{\rm{M}}$ \\
 (M$_{\odot}$)&   &(L$_{\odot}$)&(M$_{\odot}$)&& (yrs)    &    &   &(M$_{\odot}$)&& (yrs) & & \\ 
\hline
\multicolumn{13}{c}{Z = 0.02}  \\
\hline
7.0   &0.327 &4.35E+04&4.45E-02& 38&1.11E+05&0.405&0.00& 5.58E-02& 45 &1.41E+05 &2.191 &0.06  \\
7.0M  &0.327 &5.73E+04&5.00E-01&321&1.00E+06&0.770&0.00& 5.18E-01&333 &1.04E+06 &3.015 &0.02  \\
7.5   &0.327 &5.00E+04&3.86E-02& 45&9.14E+04&0.416&0.00& 5.37E-02& 58 &1.27E+05 &1.991 &0.11  \\
7.5M  &0.327 &6.26E+04&2.75E-01&264&5.09E+05&0.750&0.00& 2.90E-01&278 &5.41E+05 &2.441 &0.04  \\
8.0   &0.337 &6.00E+04&3.82E-02& 62&7.78E+04&0.417&0.00& 5.26E-02& 78 &1.08E+05 &2.004 &0.09  \\
8.0M  &0.337 &6.93E+04&1.31E-01&193&2.39E+05&0.555&0.00& 1.47E-01&214 &2.71E+05 &2.956 &0.06  \\
8.5   &0.344 &7.36E+04&3.64E-02&110&6.60E+04&0.408&0.00& 5.26E-02&145 &9.63E+04 &2.177 &0.10  \\
8.5M  &0.341 &7.90E+04&4.75E-02&149&8.63E+04&0.385&0.00& 6.53E-02&189 &1.18E+05 &2.259 &0.09  \\
8.5B95&0.344 &6.97E+04&4.34E-03& 29&1.16E+04&0.150&0.00& 4.90E-03& 31 &1.28E+04 &0.413 &0.00 \\
8.5R75&0.345 &7.68E+04&1.35E-01&381&2.26E+05&0.740&0.00& 1.95E-01&555 &3.35E+05 &6.115 &0.27 \\
8.5VL05&0.345&7.34E+04&5.66E-02&165&9.87E+04&0.457&0.00& 7.82E-02&219 &1.37E+05 &2.709 &0.12 \\
8.5$\alpha$1.5&0.345&1.23E+05&2.40E-02&84&4.38E+04&0.157&0.00&3.48E-02&112&6.26E+04&2.100&0.06\\
8.5$\alpha$2 &0.345 &1.75E+05&1.62E-02&65&3.06E+04&0.134&0.00&2.25E-02&83 &4.16E+04&1.304&0.04 \\
9.0   &0.472 &8.67E+04&3.53E-02&221&6.39E+04&0.230&0.00&4.98E-02&297&9.20E+04&1.797&0.05\\
\hline \multicolumn{13}{c}{Z = 0.008}  \\ \hline
6.5   &0.322 &4.90E+04&1.07E-01& 63&2.58E+05&1.388&0.04&1.17E-01& 68 & 2.84E+05 & 5.189  &0.13\\
6.5M  &0.323 &6.38E+04&4.88E-01&283&1.04E+06&1.082&0.01&4.98E-01&288 & 1.06E+06 & 6.068  &0.03\\
7.0   &0.324 &5.58E+04&6.38E-02& 56&1.49E+05&1.180&0.04&7.28E-02& 63 & 1.74E+05 & 4.200  &0.18\\
7.0M  &0.325 &7.26E+04&4.18E-01&356&8.13E+05&1.709&0.03&4.33E-01&368 & 8.40E+05 & 8.038  &0.06\\
7.5   &0.344 &6.48E+04&4.18E-02& 58&8.95E+04&1.067&0.03&5.25E-02&67  & 1.12E+05 & 4.302  &0.22\\
7.5M  &0.343 &8.23E+04&2.78E-01&465&6.62E+05&1.494&0.02&2.87E-01&480 & 6.85E+05 & 6.270  &0.06\\
8.0   &0.372 &7.69E+04&3.44E-02& 93&6.60E+04&1.030&0.02&4.46E-02&114 & 8.74E+04 & 4.662  &0.26\\
8.0M  &0.364 &9.03E+04&2.00E-01&690&3.71E+05&1.482&0.02&2.13E-01&736 & 3.97E+05 & 7.197  &0.09\\
8.5   &0.707 &9.35E+04&3.11E-02&185&5.84E+04&0.718&0.00&4.17E-02&238 & 8.07E+04 & 4.014  &0.22\\
8.5M  &0.710 &1.01E+05&8.06E-02&601&1.54E+05&0.675&0.00&9.70E-02&704 & 1.85E+05 & 5.639  &0.14\\
\hline
\multicolumn{13}{c}{Z = 0.004}  \\
\hline
6.5   &0.312 &6.00E+04&7.79E-02& 68&1.73E+05&3.222&0.11&8.61E-02 &73 & 1.91E+05 & 10.223 &0.19\\
6.5M  &0.318 &7.06E+04&2.08E-01&170&4.28E+05&1.913&0.08&2.16E-01&177 & 4.50E+05 & 8.298  &0.12\\
7.0   &0.352 &6.98E+04&4.32E-02& 65&9.33E+04&2.437&0.16&4.97E-02&73  & 1.09E+05 & 8.429  &0.28\\
7.0M  &0.352 &8.22E+04&1.70E-01&238&3.27E+05&5.560&0.14&1.76E-01&245 & 3.41E+05 & 13.878 &0.18\\
7.5   &0.449 &8.29E+04&3.11E-02&104&6.16E+04&2.140&0.14&4.02E-02&125 & 7.98E+04 & 8.921  &0.34\\
7.5M  &0.402 &9.79E+04&1.32E-01&437&2.45E+05&4.299&0.14&1.41E-01&460 & 2.61E+05 & 12.103 &0.19\\
8.0   &1.063 &1.00E+05&2.74E-02&161&5.35E+04&0.440&0.02&4.20E-02&239 & 8.27E+04 & 6.266  &0.27\\
8.0M  &1.024 &1.10E+05&9.75E-02&955&1.89E+05&1.042&0.01&1.10E-01&1101& 2.15E+05 & 9.295  &0.13\\
\hline 
\end{tabular}
\medskip\\
\end{center}
\end{table*}

\begin{table*}
\begin{center}
\setlength{\tabcolsep}{3.5pt} 
\caption{Standard model yields (VW-93) and VW-M in [X/Fe]. Bolded values are for the VW-M models.}
\label{recapitulation}
\begin{tabular}{lcccccccc} \hline
$M_{\rm{ini}}$ & [C/Fe] & [N/Fe] & [O/Fe] & [F/Fe] & [Ne/Fe] & [Na/Fe] &[Mg/Fe] &[Al/Fe] \\
\hline \multicolumn{9}{c}{Z=0.02} \\\hline
7.0 & -0.44/\bf{-0.18}&0.83/\bf{1.48}&-0.08/\bf{-0.20}&-0.22/\bf{-0.14}&0.02/\bf{0.25}&0.26/\bf{0.35}&0.02/\bf{0.52}&0.02/\bf{0.29} \\
7.5& -0.44/\bf{-0.27}&0.82/\bf{1.30} &-0.09/\bf{-0.22} & -0.29/\bf{-0.51} & 0.01/\bf{0.11} & 0.28/\bf{0.18} & 0.02/\bf{0.32} & 0.01/\bf{0.15} \\
8.0  & -0.45/\bf{-0.39}  &0.84/\bf{1.09} &-0.10/\bf{-0.18} & -0.44/\bf{-0.78} & 0.02/\bf{0.04} & 0.28/\bf{0.15} & 0.04/\bf{0.16} & 0.01/\bf{0.06} \\
8.5& -0.45/\bf{-0.51}  &0.84/\bf{0.90} &-0.10/\bf{-0.12} & -0.60/\bf{-0.97} & 0.01/\bf{0.02} & 0.27/\bf{0.22} & 0.05/\bf{0.06} & 0.01/\bf{0.02} \\
9.0  & -0.49/\bf{n/a}  &0.88/\bf{n/a} &-0.09/\bf{n/a} & -0.63/\bf{n/a} & 0.02/\bf{n/a} & 0.27/\bf{n/a} & 0.05/\bf{n/a} & 0.01/\bf{n/a} \\
\hline \multicolumn{9}{c}{Z=0.008} \\ \hline
6.5 & -0.09/\bf{-0.02} & 1.27/\bf{1.87} &-0.15/\bf{-0.28} & -0.18/\bf{-0.31} & 0.08/\bf{0.43} & 0.23/\bf{0.48} & 0.10/\bf{0.73} & 0.05/\bf{0.48} \\
7.0   & -0.19/\bf{0.04} & 1.12/\bf{1.79} &-0.15/\bf{-0.33} & -0.43/\bf{-0.49} & 0.04/\bf{0.29} & 0.21/\bf{0.24} & 0.06/\bf{0.67} & 0.03/\bf{0.45} \\
7.5 & -0.23/\bf{-0.03} & 1.03/\bf{1.73} &-0.12/\bf{-0.37} & -0.68/\bf{-0.82} & 0.02/\bf{0.18} & 0.22/\bf{0.00} & 0.05/\bf{0.62} & 0.02/\bf{0.46} \\
8.0   & -0.26/\bf{-0.15} & 0.99/\bf{1.53} &-0.11/\bf{-0.38} & -0.81/\bf{-1.07} & 0.02/\bf{0.07} & 0.18/\bf{-0.47} & 0.06/\bf{0.36} & 0.02/\bf{0.36} \\
8.5 & -0.25/\bf{-0.33} & 1.07/\bf{1.30} &-0.08/\bf{-0.24} & -0.78/\bf{-1.24} & 0.03/\bf{0.05} & 0.19/\bf{-0.47} & 0.07/\bf{0.15} & 0.02/\bf{0.15} \\
\hline \multicolumn{9}{c}{Z=0.004} \\ \hline
6.5 & 0.10/\bf{0.05} & 1.42/\bf{1.80} &-0.25/\bf{-0.36} & -0.36/\bf{-0.67} & 0.07/\bf{0.20} & 0.11/\bf{0.04} & 0.12/\bf{0.42} & 0.08/\bf{0.31} \\
7.0   & -0.04/\bf{0.23} & 1.23/\bf{1.73} &-0.21/\bf{-0.42} & -0.65/\bf{-0.60} & 0.03/\bf{0.14} & 0.07/\bf{-0.21} & 0.07/\bf{0.41} & 0.06/\bf{0.36} \\
7.5 & -0.06/\bf{0.13} & 1.16/\bf{1.66} &-0.16/\bf{-0.46} & -0.95/\bf{-1.10} & 0.03/\bf{0.08} & 0.03/\bf{-0.51} & 0.07/\bf{0.34} & 0.06/\bf{0.41} \\
8.0   & -0.24/\bf{-0.14} & 1.26/\bf{1.63} &-0.05/\bf{-0.38} & -0.89/\bf{-1.37} & 0.14/\bf{0.07} & 0.04/\bf{-0.66} & 0.06/\bf{0.24} & 0.05/\bf{0.40} \\
\hline
\end{tabular}
\medskip\\
\end{center}
\end{table*}

\end{document}